\documentstyle[times,pramana,epsf,floats,epsfig]{ias}

\def\gs{\mathrel{
   \rlap{\raise 0.511ex \hbox{$>$}}{\lower 0.511ex \hbox{$\sim$}}}}
\def\ls{\mathrel{
   \rlap{\raise 0.511ex \hbox{$<$}}{\lower 0.511ex \hbox{$\sim$}}}}

\newcommand{\ra}{\rightarrow}
\newcommand{\bi}{\begin{itemize}}
\newcommand{\ei}{\end{itemize}}
\newcommand{\ba}{\begin{array}{c}}
\newcommand{\baz}{\begin{array}{cc}}
\newcommand{\bad}{\begin{array}{ccc}}
\newcommand{\bav}{\begin{array}{cccc}}
\newcommand{\baf}{\begin{array}{ccccc}}
\newcommand{\bea}{\begin{equation} \begin{array}{c}}
\newcommand{\eea}{ \end{array} \end{equation}}
\newcommand{\ea}{\end{array}}
\newcommand{\D}{\displaystyle}
\newcommand{\bc}{\begin{center}}
\newcommand{\ec}{\end{center}}

\newcommand{\be}{\begin{equation}}
\newcommand{\ee}{\end{equation}}
\newcommand{\dms}{\mbox{$\Delta m^2_{\odot}$}}
\newcommand{\dma}{\mbox{$\Delta m^2_{\rm A}$}}
\newcommand{\meff}{\mbox{$\langle m \rangle$}}

\begin{document}
\mark{{The see-saw mechanism}{W Rodejohann}}
\title{The see-saw mechanism: neutrino mixing, leptogenesis and 
lepton flavor violation}

\author{WERNER RODEJOHANN}
\address{Max-Planck-Institut f\"ur Kernphysik, Postfach 103980, D-69029 
Heidelberg, Germany}
\keywords{neutrinos, leptogenesis, lepton flavor violation}
%\pacs{2.0}
\abstract{
The see-saw mechanism to generate small neutrino masses is 
reviewed. After summarizing our current knowledge about the low energy 
neutrino mass matrix we consider reconstructing  
the see-saw mechanism. Low energy neutrino physics is not 
sufficient to reconstruct see-saw, a feature which we refer to 
as ``see-saw degeneracy''. 
Indirect tests of see-saw are leptogenesis and 
lepton flavor violation in supersymmetric scenarios, which together 
with neutrino mass and mixing define the framework of 
see-saw phenomenology. Several examples are given, both 
phenomenological and GUT-related. 
Variants of the see-saw mechanism like the type II  
or triplet see-saw are also discussed. 
In particular, we compare many general aspects 
regarding the dependence of LFV 
on low energy neutrino parameters in the extreme 
cases of a dominating 
conventional see-saw term or a dominating triplet term. 
For instance, the absence of $\mu \ra e \gamma$ or $\tau \ra e \gamma$ 
in the pure triplet case means that CP is conserved in neutrino 
oscillations. 
Scanning models, we also find that among the decays  
$\mu \ra e \gamma$, $\tau \ra e \gamma$ and $\tau \ra \mu \gamma$ 
the latter one has the largest branching ratio 
in (i) $SO(10)$ type I see-saw models and in (ii) scenarios 
in which the triplet term dominates in the neutrino mass matrix.

}

\maketitle

\section{Introduction: the Neutrino Mass Matrix}
Non-trivial lepton mixing in the form of neutrino oscillations proves 
that neutrinos are massive and that the Standard Model (SM) of 
elementary particles is incomplete. At low energy, all 
phenomenology can be explained by the neutrino mass matrix 
\cite{APS}
\be \label{eq:mnu}
m_\nu = U \, m_\nu^{\rm diag} \, U^T~,
\ee
where $m_\nu^{\rm diag} = {\rm diag}(m_1, m_2, m_3)$  
contains the individual neutrino masses. 
In the basis in which the charged lepton mass 
matrix is real and diagonal 
$U$ is the Pontecorvo-Maki-Nakagawa-Sakata (PMNS) mixing matrix. 
We will work in this very basis throughout the text, 
otherwise the relation 
$U = U_\ell^\dagger \, U_\nu$ holds, where $U_\nu$ 
diagonalizes the neutrino mass matrix and 
$m_\ell \, m_\ell^\dagger = 
U_\ell \, (m_\ell^{\rm diag})^2 \, U_\ell^\dagger$. 
The PMNS matrix can explicitly be parameterized as 
\bea \label{eq:Upara}
\hspace{-1.4cm}U = 
R_{23}(\theta_{23})\,U^{\dagger}_{\delta}\,
R_{13}(\theta_{13})\,U_{\delta}\,R_{12}(\theta_{12}) \, P\\ \hspace{-1.4cm}
=     
\left( \bad 
c_{12} \, c_{13}
& s_{12}  \, c_{13} 
& s_{13}  \, e^{-i \delta} \\ 
-s_{12}   \, c_{23} 
- c_{12}   \, s_{23}   
\, s_{13}   \, e^{i \delta} 
& c_{12} \, c_{23}  - s_{12}
  \,   s_{23}  \, s_{13}  \,  e^{i \delta}
& s_{23} \,  c_{13}  \\ 
s_{12}  \, s_{23} - c_{12}  
\,   c_{23}  \, s_{13}  \, e^{i \delta} & 
-c_{12}  \, s_{23} - s_{12}  
\, c_{23}   \, s_{13} \,    e^{i \delta}
& c_{23}   \,  c_{13}  \\ 
               \ea   \right) P~.
\eea
Here 
$R_{ij}(\theta_{ij})$ is a rotation with angle $\theta_{ij}$ 
around the $ij$-axis, 
$U_{\delta}  = {\rm diag}(e^{i\delta/2},\,1,\,e^{-i\delta/2})$, 
$c_{ij} = \cos \theta_{ij}$, $s_{ij} = \sin \theta_{ij}$ 
and $P = {\rm diag}(1, \, e^{i \alpha}, \, e^{i (\beta + \delta)})$ 
contains the Majorana phases. 
CP violation in neutrino oscillation 
experiments can be described through a rephasing (Jarlskog) 
invariant quantity given by \cite{jcp} 
\bea \label{eq:jcp0}
J_{\rm CP} = {\rm Im} 
\left\{ U_{e1} \, U_{\mu 2} \, U_{e 2}^\ast \, U_{\mu 1}^\ast \right\} 
= \frac{\D {\rm Im} \left\{ h_{12} \, h_{23} \, h_{31} \right\} }
{\D \Delta m^2_{21} \, \Delta m^2_{31} \, \Delta m^2_{32}~}~, 
\eea
where $h = m_\nu \, m_\nu^\dagger$. 
With the parameterization of eq.~(\ref{eq:Upara}) one has  
$J_{\rm CP} = \frac{1}{8} \, \sin 2 \theta_{12}\,  
\sin 2 \theta_{23}\, \sin 2 \theta_{13}\, 
\cos\theta_{13}\, \sin\delta$. 
All in all, nine physical parameters are present in $m_\nu$. 
Neutrino physics deals with explaining and determining them.\\ 

To very good precision the 
angles $\theta_{12}$, $\theta_{23}$ and $\theta_{13}$ correspond to 
the mixing angles in solar (and long-baseline reactor), atmospheric 
(and long-baseline accelerator) and short-baseline reactor neutrino 
experiments, respectively. 
The analyses of neutrino experiments revealed the 
following best-fit values 
and $3\sigma$ ranges of the oscillation 
parameters \cite{newdata}: 
\begin{eqnarray}
\dms \equiv m_2^2 - m_1^2 &=& 
\left(7.67^{+0.67}_{-0.61}\right) 
\cdot 10^{-5} ~{\rm eV}^2~,\nonumber\\
\sin^2 \theta_{12} &=& 0.32^{+0.08}_{-0.06} ~,\nonumber\\
\dma \equiv \left| m_3^2 - m_1^2 \right|&=&  
\left\{ \baz 
\left(2.46^{+0.47}_{-0.42}\right) 
\cdot 10^{-3} ~{\rm eV}^2 & \mbox{for } m_3^2 > m_1^2\\[0.1cm]
\left(2.37^{+0.43}_{-0.46}\right) 
\cdot 10^{-3} ~{\rm eV}^2 & \mbox{for } m_3^2 < m_1^2
\ea \right. ~,\\
\sin^2\theta_{23} &=& 0.45^{+0.20}_{-0.13} ~,\nonumber\\
|U_{e3}|^2  &=&0^{+0.050}_{-0.000} ~.\nonumber
\end{eqnarray}
Depending on the sign of $m_3^2 - m_1^2$, 
the neutrino masses are normally 
or inversely ordered: 
\[ %be \nonumber
%\label{eq:masses}
\bad
\text{normal:}  & \text{with} 
& m_2 = \sqrt{m_1^{2} + \dms} ~;~~~~~~  
m_3 = \sqrt{m_1^{2} + \dma} ~,\\%[0.3cm]
\text{inverted:} &  \text{with} & 
m_2 = \sqrt{m_3^{2} + \dms + \dma} ~;~~~~ 
m_1 = \sqrt{m_3^{2} + \dma}\,.
\ea
\] %ee
The overall scale of neutrino masses is not known, except 
for the upper limit of order 1 eV coming from direct mass search 
experiments and cosmology. The hierarchy of the light neutrinos, 
at least between the two heaviest ones, is 
moderate: 
\be \label{eq:hie}
\hspace{-1.4cm}\mbox{normal: } 
\frac{m_2}{m_3} \ge \sqrt{\frac{\dms}{\dma}} \simeq 0.17~;~~
\mbox{inverted: } 
\frac{m_1}{m_2} \gs 1 - \frac 12 \, \frac \dms\dma \simeq 0.98~.
\ee 
These numbers should be compared, e.g., with $m_e/m_\mu \simeq 1/200$. 

The current data for the mixing angles 
can accurately be described by tri-bimaximal 
mixing \cite{tbm}:
\be \label{eq:tbm}
\hspace{-1.4cm}U \simeq U_{\rm TBM} =  
R_{23} (-\pi/4) \, R_{12}(\sin^{-1} 1/\sqrt{3}) \, P
= \left(
\bad
\sqrt{\frac{2}{3}} & \sqrt{\frac{1}{3}}  & 0 \\
-\sqrt{\frac{1}{6}} & \sqrt{\frac{1}{3}} & 
-\sqrt{\frac{1}{2}}  \\
-\sqrt{\frac{1}{6}} 
& \sqrt{\frac{1}{3}} &  \sqrt{\frac{1}{2}}
\ea \right) P~.  
\ee
Any parameterization of the PMNS matrix must build 
upon tri-bimaximal mixing. 
A recent proposal to phenomenologically 
take into account (expected) deviations from tri-bimaximal 
mixing is the triminimal parameterization \cite{PRW}
\bea \label{eq:tmin}
U_{\rm Tmin} = 
R_{23} (-\pi/4) \, 
\,U_\epsilon(\epsilon_{23};\epsilon_{13},\delta;\epsilon_{12})
R_{12}(\sin^{-1} 1/\sqrt{3}) \, P \\ 
= 
R_{23} (-\pi/4) \, 
R_{23}(\epsilon_{23}) \, U_\delta^\dagger\, 
R_{13}(\epsilon_{13}) \,U_\delta \, R_{12}(\epsilon_{12}) \, 
R_{12}(\sin^{-1} 1/\sqrt{3}) \, P~.
\eea
In contrast to other parameterizations of the PMNS matrix the 
triminimal one has the virtue that 
each $\epsilon_{ij}$ is directly interpretable as the deviation of, 
and only of, the associated 
$\theta_{23}$, $\theta_{13}$, or $\theta_{12}$ from its  
tri-bimaximal value. 
If it turns out that one of the deviations from 
tri-bimaximal mixing is sizable, this parametrization 
can treat that case more accurately.  
One easily finds that 
\begin{eqnarray} \label{eq:obsII} \D 
\sin^2 \theta_{12} &=& \frac 13 \left( \cos \epsilon_{12} + 
\sqrt{2} \, \sin \epsilon_{12} \right)^2  %\\ & \simeq  & \nonumber 
\simeq \frac 13 + \frac{2\sqrt{2}}{3} \, \epsilon_{12} 
 + \frac{1}{3} \, \epsilon_{12}^2 ~, \nonumber \\%[0.2cm]
\D
\sin^2 \theta_{23} &=& \frac 12 - \sin \epsilon_{23} \, \cos \epsilon_{23} 
\simeq  \frac 12 - \epsilon_{23} ~,\\%[0.2cm]
\D U_{e3} &=& \sin \epsilon_{13} \, e^{-i \delta} \, 
e^{i (\beta + \delta)}
\nonumber  ~.
\end{eqnarray}
One sees in the above expressions that the triminimal parametrization 
maintains the simple parametrization of $U_{e3}$.
Being a 3-flavor quantity, 
$J_{\rm CP}$ depends on all three $\epsilon_{jk}$. Its expansion 
reads 
\begin{eqnarray} 
\D
J_{\rm CP} &=& -\frac{\sin \delta}{24} \, 
\cos 2 \epsilon_{23} \, \sin 2 \epsilon_{13} 
\, \cos \epsilon_{13} \left(2\sqrt{2} \, \cos 2  \epsilon_{12} 
+  \sin 2\epsilon_{12}\right) \\ \nonumber 
& \simeq &   \frac{-1}{3 \sqrt{2}}\, 
\left(1+\frac{\epsilon_{12}}{\sqrt{2}}\right) 
 \epsilon_{13} \, \sin \delta\,.
\end{eqnarray}
Tri-bimaximal mixing is a special case of $\mu$--$\tau$ symmetry, which 
implies $\theta_{23} = -\pi/4$ and $\theta_{13}=0$. The mass matrices 
for $\mu$--$\tau$ symmetry and for tri-bimaximal mixing are   
\bea \label{eq:mnumutau}
%\hspace{-1.4cm}
(m_\nu)^{\mu\mbox{--}\tau} = 
\left( 
\bad 
A & B & B \\
\cdot & D & E \\ 
\cdot & \cdot & D 
\ea
\right) ~,~\\
(m_\nu)^{\rm TBM} = 
\left(
\bad 
\tilde A & \tilde B & \tilde B  \\
\cdot & \frac{1}{2} (\tilde A + \tilde B + \tilde D) 
& \frac{1}{2} (\tilde A + \tilde B - \tilde D)\\
\cdot & \cdot & \frac{1}{2} (\tilde A + \tilde B + \tilde D)
\ea 
\right)~,
\eea
where the $\stackrel{(\sim)}{A}, 
\stackrel{(\sim)}{B},\stackrel{(\sim)}{D},E$ 
are functions of the neutrino masses, Majorana phases, and 
in case of $\mu$--$\tau$ symmetry, $\theta_{12}$. 
Writing the 
tri-bimaximal neutrino 
mass matrix in terms of matrices multiplied 
with the individual neutrino masses gives: 
\be \label{eq:mnutbm}
\hspace{-1.84cm}(m_\nu)^{\rm TBM} = \frac{m_1}{6} \, 
\left( 
\bad 
4 & -2 & -2 \\
\cdot & 1 & 1 \\
\cdot & \cdot & 1 
\ea 
\right) 
+ \frac{m_2 \, e^{2i \alpha}}{3} \, 
\left( 
\bad 
1 & 1 & 1 \\
\cdot & 1 & 1 \\
\cdot & \cdot & 1 
\ea 
\right) + 
\frac{m_3 \, e^{2i \beta}}{2} \, 
\left( 
\bad 
0 & 0 & 0 \\
\cdot & 1 & -1 \\
\cdot & \cdot & 1 
\ea 
\right)\,.
\ee
Interestingly, the state with mass $m_2$ is democratic, i.e., couples 
with equal strength to all flavors.

\begin{table}[ht]
\caption{\label{tab:mnu}Candidate mass matrices with an eigenvector 
$(0,\,-1/\sqrt{2}, \,1/\sqrt{2})^T$, 
the corresponding eigenvalue with the resulting mass ordering and 
the $U(1)$ leading to it. NH is the normal 
hierarchy $(m_3^2 \simeq \dma \gg m_2^2 \simeq \dms \gg m_1^2)$, 
IH the inverted hierarchy $(m_2^2 \simeq m_1^2 \simeq \dma \gg 
m_3^2)$ and QD denote quasi-degenerate neutrinos 
$(m_3^2 \simeq m_2^2 \simeq m_1^2 \gg \dma, \dms)$.}
\begin{tabular}{ccc} 
$m_\nu$ & eigenvalue & $U(1)$ \\ \hline  
$\D \sqrt{\frac{\dma}{4}}
\left( 
\bad
0 & 0 & 0 \\
\cdot & 1 & -1 \\
\cdot & \cdot & 1
\ea
\right)$  & 
$\sqrt{\dma} \Rightarrow$ NH  & 
$L_e$ \\ \hline
$ \D \sqrt{\frac{\dma}{2}} 
\left( 
\bad
0 & 1 & 1 \\
\cdot & 0 & 0 \\
\cdot & \cdot & 0 
\ea
\right)$ 
& $0 \Rightarrow$ IH & 
$L_e - L_\mu - L_\tau$ \\ \hline
$ m_0 
\left( 
\bad
1 & 0 & 0 \\
\cdot & 0 & 1 \\
\cdot & \cdot & 0 
\ea
\right) $ 
& $-m_0 \Rightarrow$ QD 
&  $L_\mu - L_\tau$  \\ 
\end{tabular}
\end{table}
The (approximate) $\mu$--$\tau$ symmetry indicates 
that the neutrino mass matrix has an 
eigenvector of the form $(0,\,-1/\sqrt{2}, \,1/\sqrt{2})^T$. 
This property is fulfilled by three simple and frequently used candidate 
mass matrices, summarized in table \ref{tab:mnu}. 
The candidates can be interpreted as a consequence of a conserved 
$U(1)$ lepton charge. The conservation is only 
approximate, moderate breaking is necessary to 
obtain full agreement with data. We note here 
that the $U(1)$ symmetry allows strictly speaking only for order 
one terms in the non-zero entries, which are in general 
not equal to each other. 
Only $L_\mu - L_\tau$ is automatically $\mu$--$\tau$ symmetric 
\cite{CR}. 

Another proposal for the mass matrix is introduced by 
the requirement of ``scaling''. This denotes the property that 
the ratios of mass matrix elements $(m_\nu)_{\alpha \mu}$ and 
$(m_\nu)_{\alpha \tau}$ are independent of $\alpha$:  
\be \label{eq:SSA}
m_\nu = 
\left( 
\bad
A & B & B/c \\
\cdot & D & D/c \\
\cdot & \cdot & D/c^2
\ea
\right)\,.
\ee 
One easily finds that an inverted hierarchy is predicted and 
that $m_3 = 0$ (the rank of this matrix is two).  
Furthermore, $\theta_{13}$ is zero and the scaling factor $c$ governs 
atmospheric neutrino mixing: 
$\tan^2 \theta_{23} = 1/c^2$ \cite{SSA,SSA2}.

Leaving concrete models and Ans\"atze aside, 
we can use our current knowledge of the neutrino parameters to 
reconstruct $m_\nu$. Varying the neutrino mixing 
parameters and the Majorana phases in their allowed ranges, one can plot 
the individual mass matrix entries $|(m_\nu)_{\alpha \beta}|$ for both 
mass orderings as a function of the smallest neutrino mass \cite{MR}: 

$m_\nu = \left(
\begin{minipage}{10.480cm}
\begin{tabular}[h]{ccc}
\epsfig{file=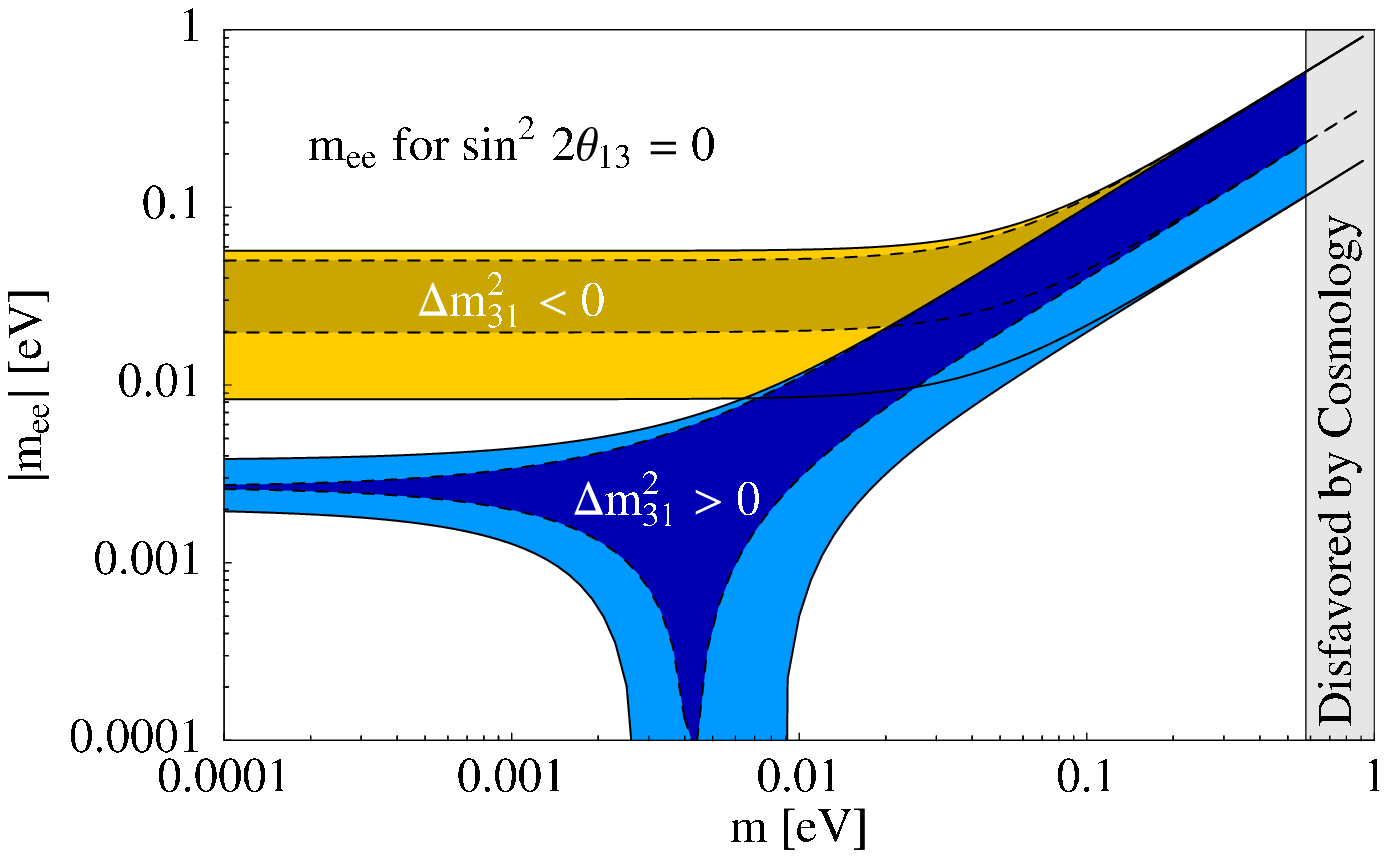,width=3.3275cm,height=2.572cm}  &
\epsfig{file=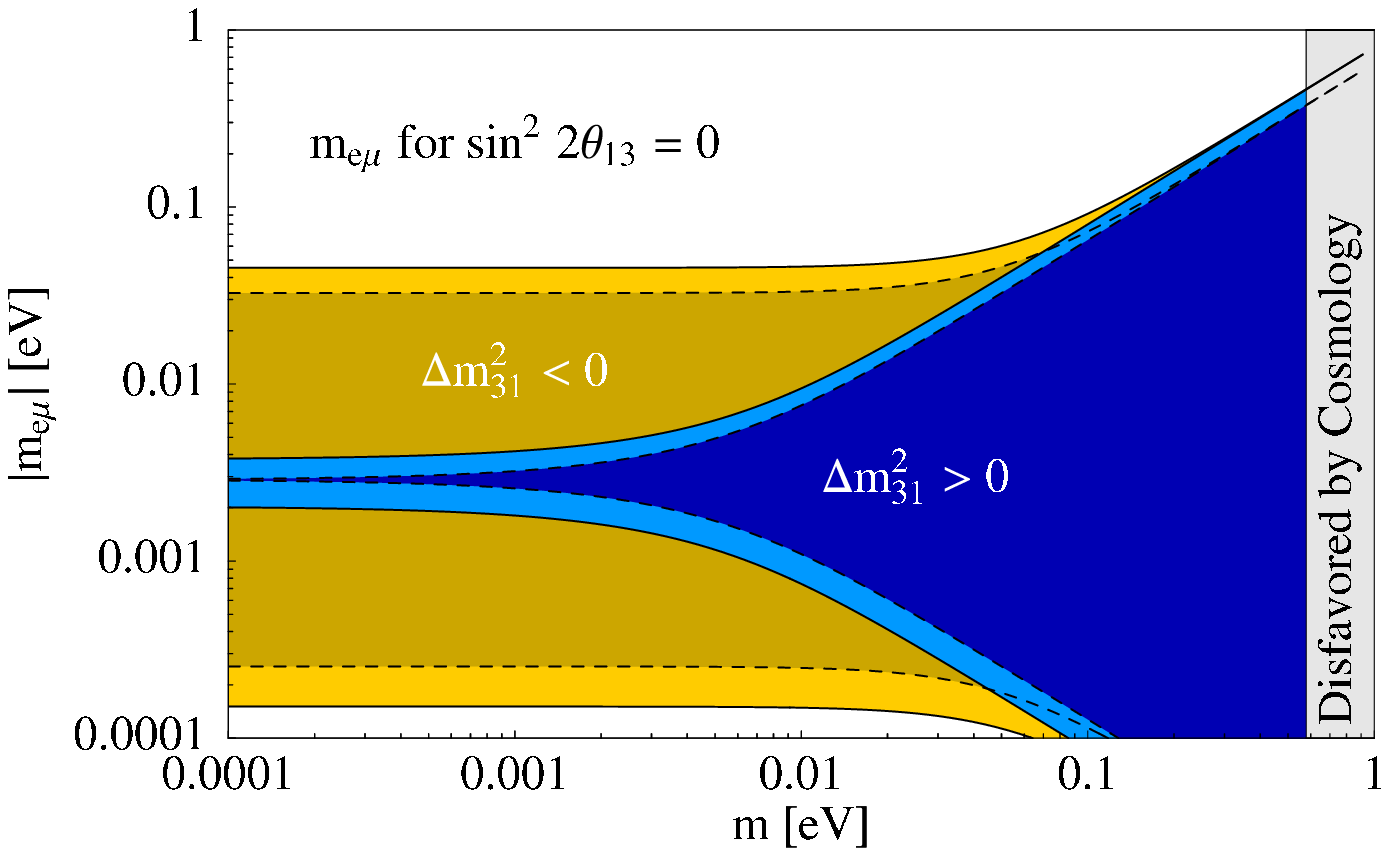,width=3.3275cm,height=2.572cm} &
\epsfig{file=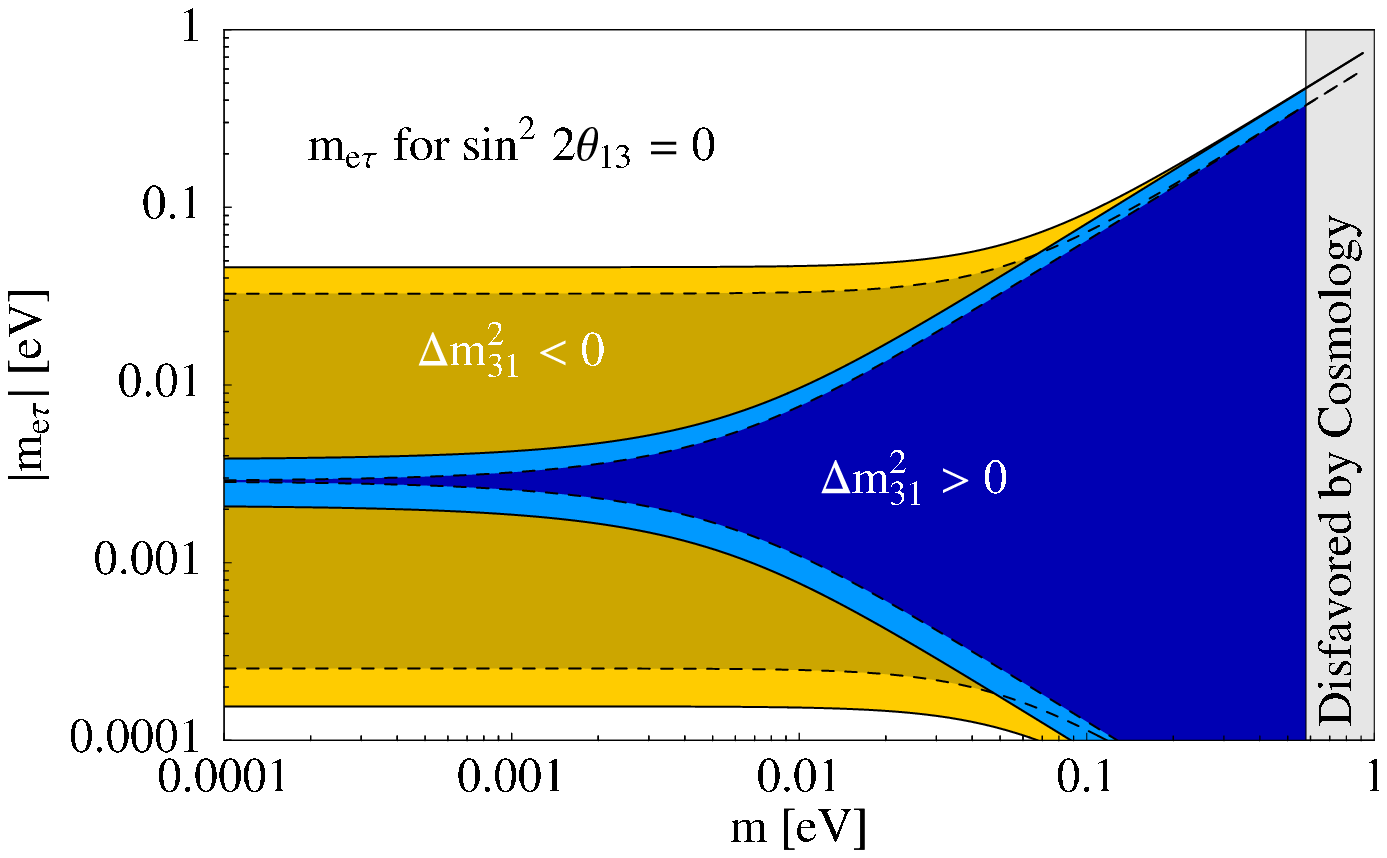,width=3.3275cm,height=2.572cm} \\
\epsfig{file=memu1.eps,width=3.3275cm,height=2.572cm} &
\epsfig{file=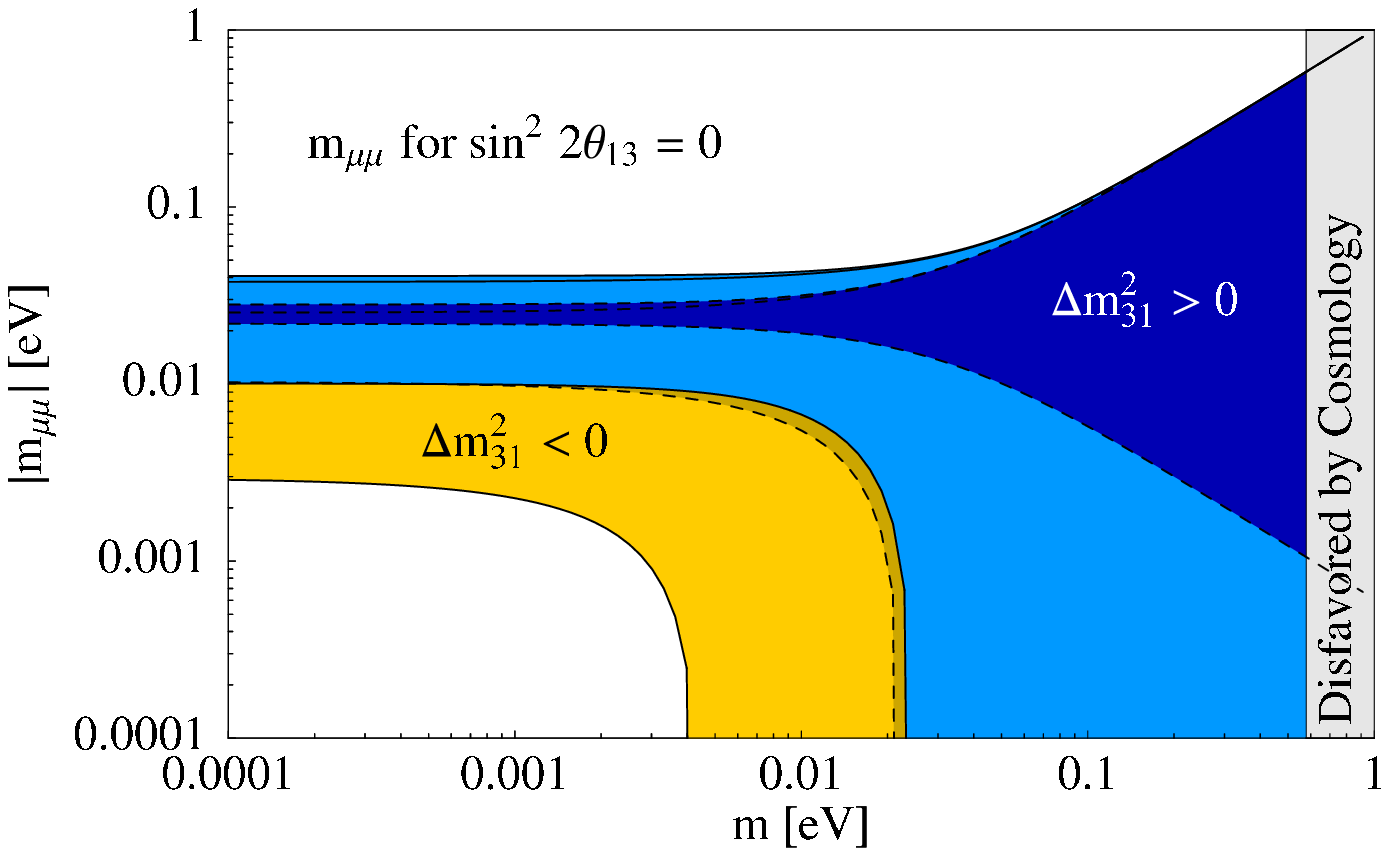,width=3.3275cm,height=2.572cm} &
\epsfig{file=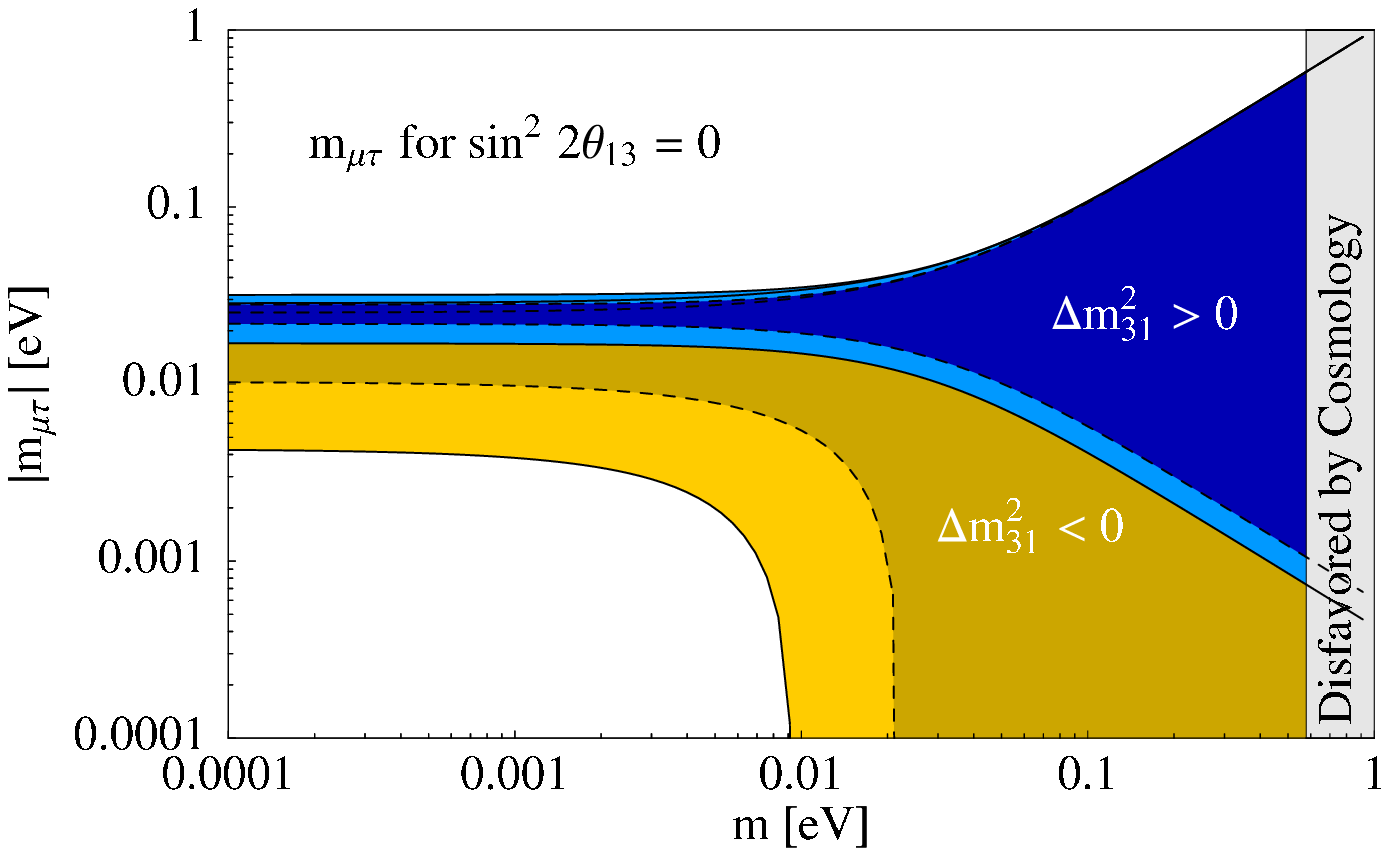,width=3.3275cm,height=2.572cm} \\
\epsfig{file=metau1.eps,width=3.3275cm,height=2.572cm} & 
\epsfig{file=mmutau1.eps,width=3.3275cm,height=2.572cm} & 
\epsfig{file=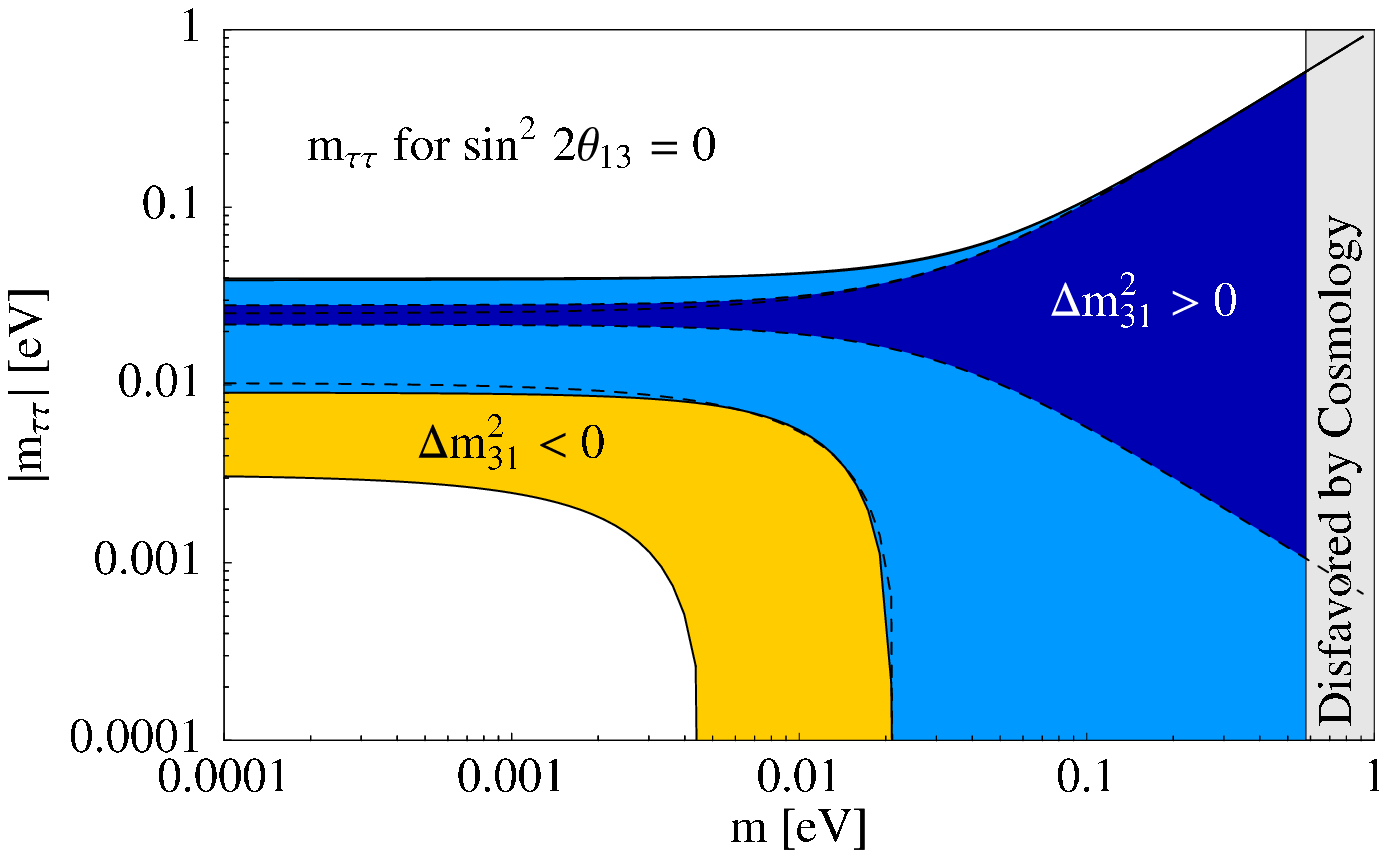,width=3.3275cm,height=2.572cm}
\end{tabular}
\end{minipage}
\right)$

The yellow (blue) bands are for the inverted (normal) mass ordering, 
and the darker areas are for the best-fit values of the oscillation 
parameters with only the Majorana phases varied. The lighter areas 
are for the current $3\sigma$ ranges of the oscillation parameters. 
The absolute value of the 
$ee$ element of the mass matrix is of course the effective mass on which 
the rate of neutrinoless double beta decay depends quadratically. It 
is a function of seven of the nine physical parameters of $m_\nu$: 
\be
|(m_\nu)_{ee}| \equiv 
\meff = \left| c_{12}^2 \, c_{13}^2 \, m_1 + s_{12}^2 \, c_{13}^2 \, 
m_2 \, e^{2i \alpha} + s_{13}^2 \, m_3 \, e^{2 i \beta} \right| ~.
\ee
Summarizing our knowledge about $m_\nu$, it is to a good 
precision given by eq.~(\ref{eq:mnumutau}) or (\ref{eq:mnutbm}), and 
can be interpreted as a result of a conserved 
lepton $U(1)$ charge. Other possible 
properties of $m_\nu$, which are perfectly compatible 
with current neutrino data, are: 
there can be zero entries in $m_\nu$, 
the maximal number is two \cite{zeros}, 
but one zero entry is also allowed \cite{MR}. In the 
(unlikely) case of neutrinos being Dirac particles, $m_\nu$ can 
have five zero entries \cite{HR}. 
The possibility of two equal elements and one zero entry 
has also been discussed \cite{hybrid}. 
Finally, the determinant \cite{det} and the 
trace \cite{trace} of $m_\nu$ can vanish. 
We refer to the given references for details of the 
resulting phenomenology.\\ 

Obviously there are many models and Ans\"atze for the neutrino mass 
matrix, simply due to the fact that many of the low energy parameters 
are currently unknown. Future precision data will sort out 
many possibilities \cite{carl0} 
and shed more light on the flavor structure 
in the lepton sector.

\section{\label{seesaw}The See-Saw Mechanism and 
its Reconstruction: the See-Saw Degeneracy} 

A most important question in this framework is about the origin 
of the neutrino mass matrix. 
One possibility to accommodate $m_\nu$ is to introduce SM singlets which 
can couple to the left-handed $\nu_L$ and the (up-type) Higgs doublet. 
Usually these singlets are right-handed neutrinos $N_{R i}$, 
and the corresponding Lagrangian is 
\bea \label{eq:LI}
{\cal L} = \frac 12 \, \overline{N_{R i}^c} \, (M_R)_{ij} \, N_{Rj}  
+  \overline{L_\alpha} \, (Y_D)_{i \alpha} \, N_{R i} \, \Phi \\ \D 
= \frac 12 \, \overline{N_{R}^c} \, M_R \, N_R + \overline{\nu_L} \, 
m_D \, N_{R}~.
\eea 
Here $m_D$ is the Dirac mass matrix expected to be related to the 
known SM masses, and $M_R$ is a (symmetric) Majorana mass matrix.  
Integrating out the heavy $N_{R i}$ ($M_R$ is not 
constrained by the electroweak scale because the $N_{R i}$ are SM 
singlets) gives the see-saw formula \cite{I} 
\be \label{eq:mnuI}
m_\nu = - m_D \, M_{R}^{-1} \, m_D^T~.
\ee
It is also known as the ``conventional'', or type I, see-saw 
formula (for other reviews on it, see \cite{revs}). 
Taking the neutrino mass scale as 
$\sqrt{\dma}$ and the scale of $m_D$ as 
$v = 174$ GeV gives $M_R \simeq 10^{15}$ GeV. We will assume 
in what follows that the see-saw particles are very heavy. 
Being a SM singlet, the first guess for $M_R$ would be 
the Planck mass, which however gives too small neutrino masses, 
though small effects from Planck scale effects may be 
present \cite{planck}. 
$M_R$ is typically also smaller than the GUT scale 
of $2 \cdot 10^{16}$ GeV, presumably $10^{15}$ GeV 
is related to the scale of $B - L$ breaking.

The main ingredient of the see-saw mechanism is the vertex 
$\overline{L_\alpha} \, (Y_D)_{i \alpha} \,  N_{R i} \, \Phi$. 
Testing this vertex is obviously crucial for testing and 
reconstructing see-saw. In this respect, note that 
the number of physical parameters in $m_D$ and $M_R$ 
is 18, six of which are phases. 
Comparing this with the number of parameters 
in $m_\nu$ we see that half of the see-saw parameters 
get lost when the heavy degrees of freedom 
are integrated out. To put it another way, we hardly know $m_\nu$ and 
we know neither $m_D$ nor $M_R$. Reconstructing the see-saw 
mechanism is therefore a formidable task \cite{davidson,ellis,PPR}, 
even more so when one notes that the see-saw scale of 
$M_R \simeq 10^{15}$ GeV is 11 orders of magnitude above the LHC 
center-of-mass energy. 
Leaving aside for now observables which 
indirectly depend on the see-saw parameters (see below), 
we have two possibilities to facilitate the reconstruction: 
(i) making assumptions about $m_D$ and/or $M_R$, and (ii) parameterize 
our ignorance:

\begin{figure}[ht]
\begin{center}
\epsfig{file=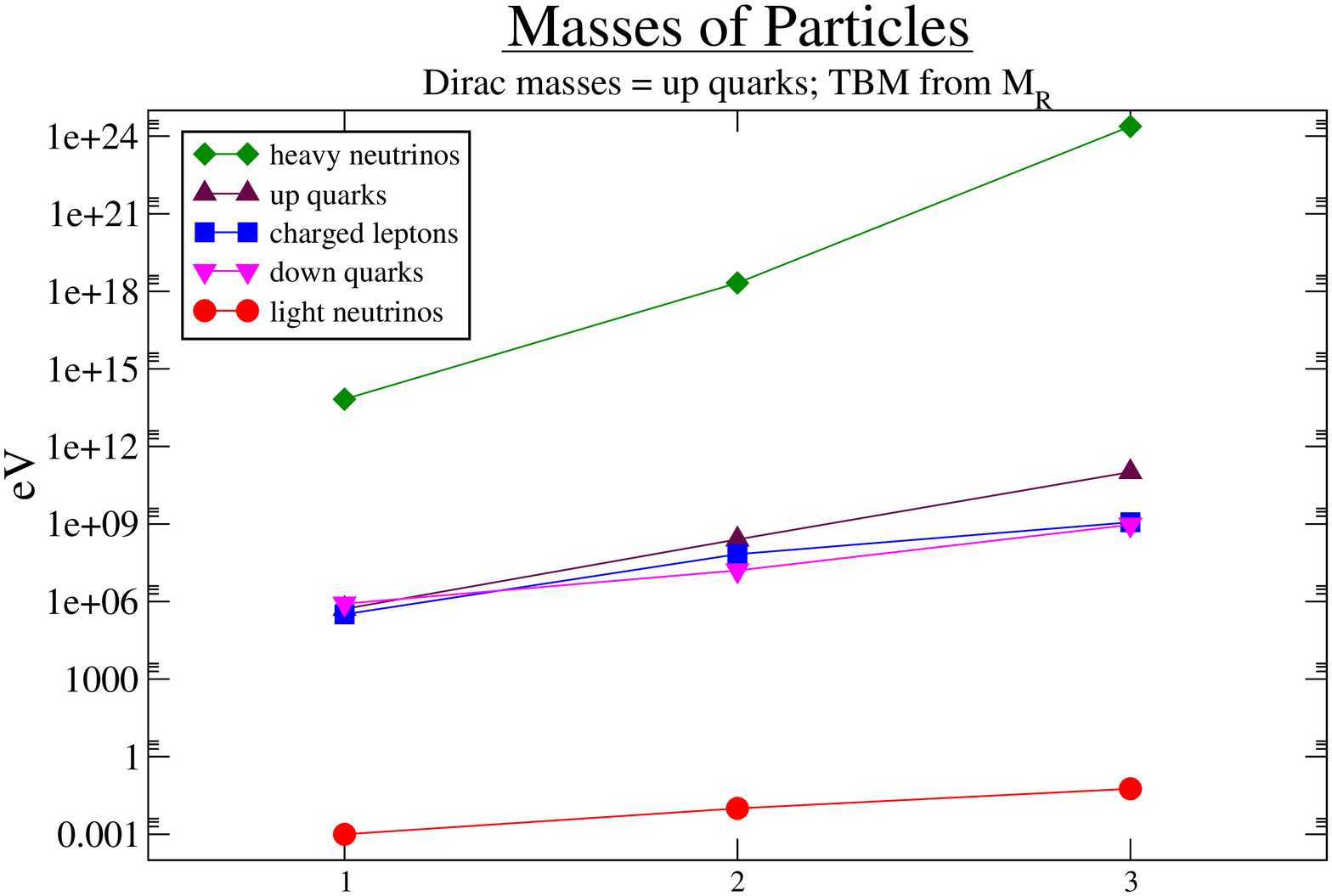,width=10cm,height=7cm} 
\caption{\label{fig:hie}Masses of the SM particles plus light and 
heavy neutrinos if the Dirac mass matrix is diag$(m_u , m_c, m_t)$ and 
$m_1 = 10^{-3}$ eV with tri-bimaximal mixing. 
We upscaled the quark masses using the value $\tan \beta = 10$.}
\end{center}
\end{figure}

{\it (i) making assumptions}

The most simple semi-realistic example is to assume that $m_D$ is the 
up-quark mass matrix. This can happen in $SO(10)$ models with a 
${\bf 10}$ Higgs representation. We can in this case 
use the see-saw formula to find
$M_R = -m_{\rm up} \, m_\nu^{-1} \, m_{\rm up}$ and diagonalize 
$M_R$ to obtain the heavy masses. 
Assuming that $m_D$ is diagonal, and inserting 
tri-bimaximal mixing and no CP phases gives \cite{branco,AFS}: 
\be \label{eq:hie2}
M_1 \simeq  3 \, \frac{2 \, m_u^2}{m_2}~,~~
M_2 \simeq \frac{2 \, m_c^2}{m_3}~,~~
M_3 \simeq \frac 13 \, \frac{m_t^2}{2 \, m_1} ~ .
\ee  
The naive see-saw expectation $m_3 \propto m_t^2$, $m_2 \propto 
m_c^2$ and $m_1 \propto m_u^2$ is completely changed due to the large 
neutrino mixing. Note that $M_1 \propto m_u^2$, $M_2 \propto m_c^2$ 
and $M_3 \propto m_t^2$, i.e., the hierarchy of the heavy neutrinos 
is the hierarchy of the up-quarks {\it squared}. This is necessary, 
in particular, to ``correct'' the strong up-quark hierarchy into the 
very mild light neutrino hierarchy, see eq.~(\ref{eq:hie}). 
Fig.~\ref{fig:hie} displays 
the masses of the three fermion families in this simple 
example. Note how the SM fermions 
are ``sandwiched'' between the light and heavy neutrinos.

\begin{figure}
\begin{center}
\epsfig{file=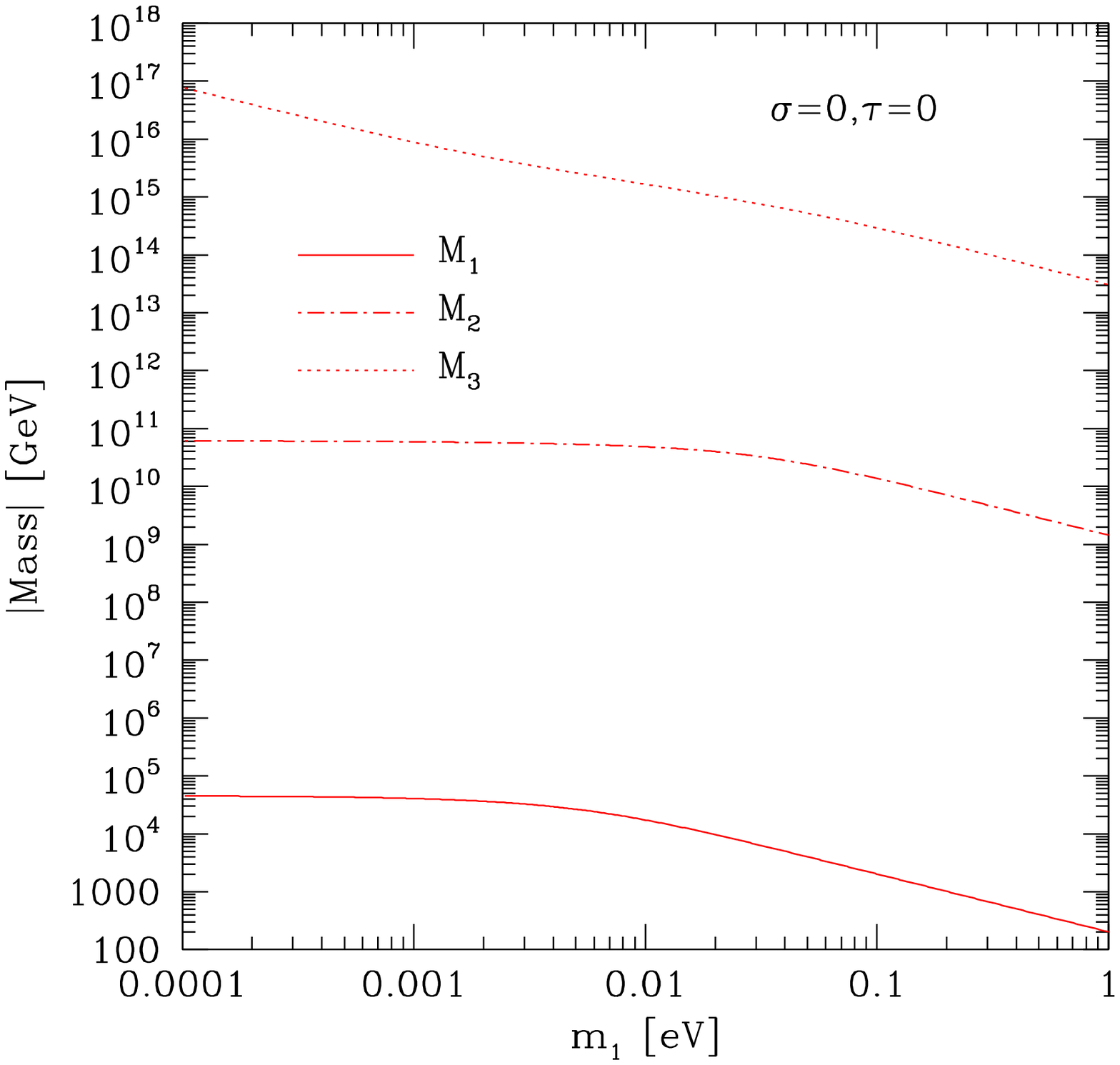,width=6.14cm,height=5.5cm}
\epsfig{file=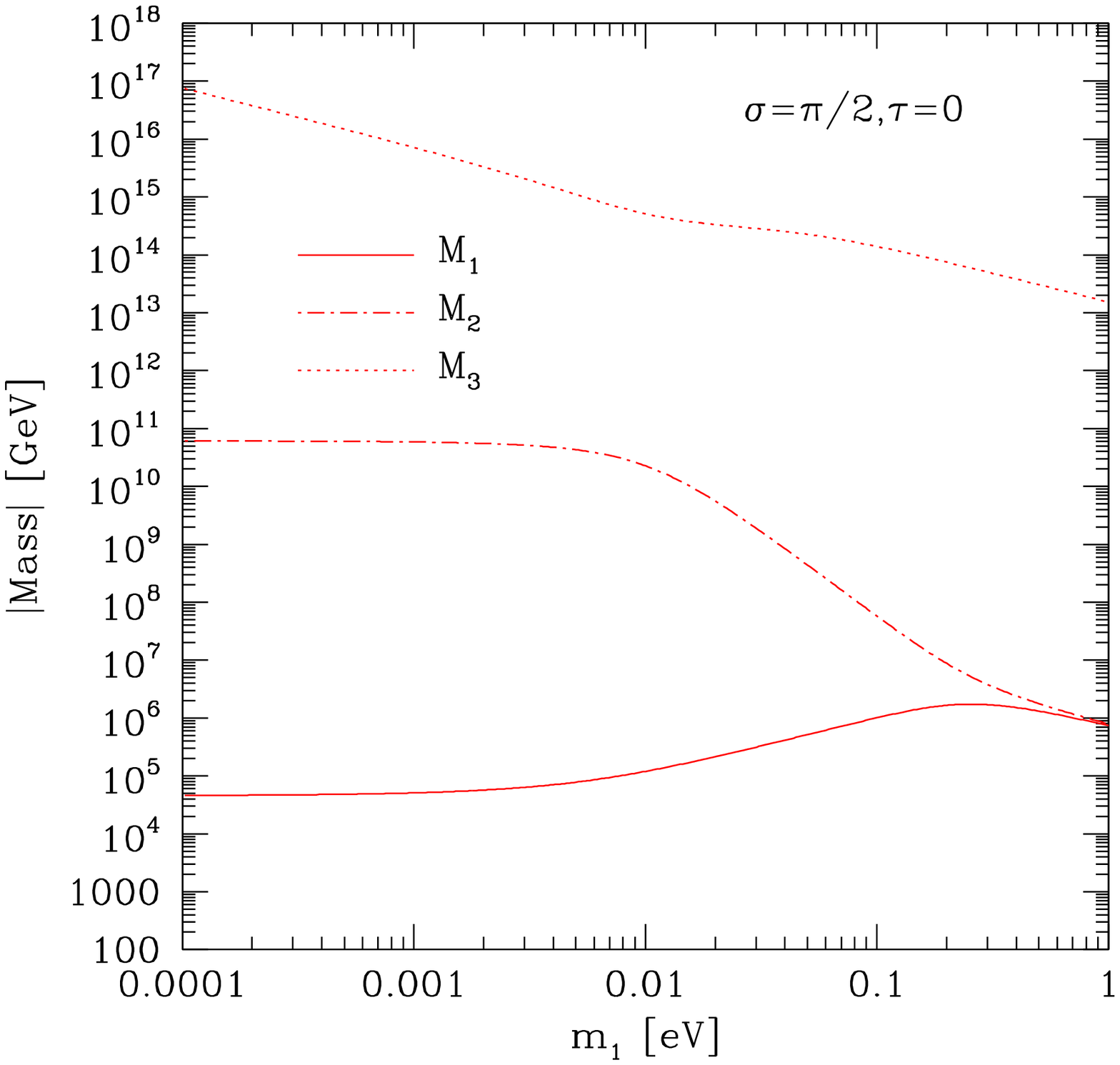,width=6.14cm,height=5.5cm} 
\caption{\label{fig:qlc}Heavy neutrino masses and non-trivial 
CP phases. Taken from \protect\cite{QLC0}.}
\end{center}
\end{figure}

The simple picture presented changes already in the presence of CP 
phases \cite{AFS}. 
Fig.~\ref{fig:qlc} (taken from ref.~\cite{QLC0}) shows the 
masses of the heavy 
neutrinos (for bimaximal neutrino mixing) for no phases and 
for one of the possible phases equal to $\pi/2$. The latter 
case can lead to degenerate heavy neutrinos. 
Even more modification occurs in realistic $SO(10)$ models. 
In table \ref{tab:so10}, 
taken from ref.~\cite{rabi_chin}, predictions for the smallest neutrino 
mass of different $SO(10)$ models, which differ in their 
Higgs content and in their flavor structure, are given  
(see also table \ref{tab:so10_carl}, which is taken 
from ref.~\cite{carl}). 
The value of $M_1$ in the simple example leading to 
eq.~(\ref{eq:hie2}) was about $10^5$ GeV, obviously very different from 
the values in the table, which also differ a lot for the various 
models. 
The reason for this large spread in seemingly similar models is 
connected to 

\begin{table}
\caption{Higgs content, predicted mass $M_1$ of the
lightest right-handed neutrino and baryon
asymmetry $\eta_B$ in various $SO(10)$ models. The prediction 
for $|U_{e3}|$ is also given. 
Taken from \protect\cite{rabi_chin} and slightly modified.}
\label{tab:so10} 
\begin{tabular}{cccccc} %\hline \hline 
 & BPW \cite{BPW}  & GMN \cite{GMN} & JLM \cite{JLM} 
& DMM \cite{DMM} & AB \cite{AB} 
%\scriptsize{(modified in \cite{rabi_chin})}  
\\ \hline
Higgs 
& ${\bf 10}$, ${\bf 16}$, $\overline{\bf 16}$, ${\bf 45}$
& ${\bf 10}$, ${\bf 210}$, $\overline{\bf 126}$ 
& ${\bf 10}$, ${\bf 16}$, $\overline{\bf 16}$, ${\bf 45}$ 
& ${\bf 10}$, ${\bf 210}$, $\overline{\bf 126}$, ${\bf 120 }$ 
& ${\bf 10}$, ${\bf 16}$, $\overline{\bf 16}$, ${\bf 45}$\\  \hline
$M_1$ [GeV] & $10^{10}$ & $10^{13}$& $3.77 \cdot 10^{10}$   & $10^{13}$
&$5.4 \cdot 10^{8}$\\  \hline
$\eta_B $& $ 12 \cdot 10^{-10} \sin 2 \phi $ & $ 5 \cdot 
10^{-10}   $&  $6.2\cdot 10^{-10}$    &  $10^{-9}\sin{2\phi}$& $
2.6 \cdot 10^{-10}$   \\  \hline
$|U_{e3}|$ & $\le 0.16$ & 0.18 & $0.12 \div 0.15$ & $0.06 \div 0.11$ 
& 0.05  
\end{tabular}
\end{table}

{\it (ii) parameterizing our ignorance: the see-saw degeneracy}

The impossibility to make unambiguous statements about the see-saw 
parameters becomes very obvious when we 
parameterize our ignorance. This can be done with the 
so-called Casas-Ibarra parametrization \cite{CI}: 
\be \label{eq:CI}
m_D = i \, U \, \sqrt{m_\nu^{\rm diag}} \, R \, \sqrt{M_R}  ~.
\ee
Here $R$ is a complex and orthogonal matrix which contains the unknown 
see-saw parameters.  Usually the 
parameterization in eq.~(\ref{eq:CI}) is considered in the basis 
in which $M_R$ is real and diagonal. In the already pretty ideal 
situation in which we knew 
$m_\nu$ and $M_R$, there would be still an infinite number of allowed 
Dirac mass matrices. We will refer to this unpleasant feature as  
``see-saw degeneracy''. 
We can parameterize the parametrization of our 
ignorance by writing $R$ as 
\be \label{eq:Rpara} 
R = R_{12} \, R_{13} \, R_{23}~,
\ee
where $R_{ij}$ is a rotation around the $ij$-axis with complex angle 
$\omega_{ij} = \rho_{ij} + i \sigma_{ij}$, 
$\rho_{ij}$ and $\sigma_{ij}$ being real. Actually, this parametrization 
does not include ``reflections'' \cite{CI}, i.e., it should be 
multiplied with $\tilde{R} \equiv 
{\rm diag}(\pm 1, \pm 1, \pm 1)$ from the left, 
where $\tilde{R}$ contains an odd number of minus signs. 
However, in many cases the implied additional forms of $R$ do not 
lead to different textures in $m_D$ and the parametrization in 
eq.~(\ref{eq:Rpara}) is general enough.

\section{See-saw at work: Lepton Flavor Violation and Leptogenesis}
We conclude from the above that reconstructing see-saw 
requires more than low energy neutrino 
physics. One observable which can in principle be used is 
the baryon asymmetry of the Universe. Lepton Flavor Violation 
(LFV) in supersymmetric scenarios can also depend on 
the see-saw parameters. 
Here we will focus on the rare decays $\ell_i \ra \ell_j \gamma$, 
with $\ell_{3,2,1} = \tau, \mu, e$. 

\subsection{\label{sec:LFV}Lepton Flavor Violation} 
LFV in supersymmetric see-saw scenarios allows decays like 
$\ell_i \ra \ell_j \gamma$, triggered by off-diagonal entries in the 
slepton mass matrix $\tilde{m}_L^2$. 
The branching ratios for radiative decays of the charged leptons 
$\ell_i = e, \mu, \tau$ are \cite{LFV}
\be \label{eq:BR}
{\rm BR}(\ell_i \ra \ell_j \gamma) 
= {\rm BR}(\ell_i \ra \ell_j \, \nu 
\bar{\nu}) \, \frac{\alpha^3}{G_F^2 \, m_S^8} 
\, \left| \left( \tilde{m}_L^2 \right)_{ij} \right|^2 
\, \tan^2 \beta\, ,
\ee
where $m_S$ is a typical mass scale of SUSY particles. 
Note the normalization factors ${\rm BR}(\ell_i \ra \ell_j 
\, \nu \overline{\nu})$ in the definition of 
the branching ratios in eq.~(\ref{eq:BR}). The 
numbers are ${\rm BR}(\mu \ra e \, \nu \overline{\nu}) = 0.178$ 
and ${\rm BR}(\tau \ra \mu \, \nu \overline{\nu}) = 0.174$ \cite{PDG}, 
respectively.  Current limits 
on the branching ratios for $\ell_i \ra \ell_j \gamma$ are 
BR$(\mu \ra e \gamma) \le 1.2 \cdot 10^{-11}$ \cite{mueg_lim}, 
${\rm BR}(\tau \ra e \gamma) \le 
1.1 \cdot 10^{-7}$ \cite{teg_lim} and 
${\rm BR}(\tau \ra \mu \gamma) \le 6.8 \cdot 10^{-8}$ 
\cite{tmg_lim}. One expects to improve these bounds by 
two to three orders of magnitude for BR$(\mu \ra e \gamma)$ 
\cite{meg_fut} and by one to two orders of magnitude 
for the other branching ratios \cite{BR_fut}.

To satisfy the requirement that the LFV branching ratios 
${\rm BR}(\ell_i \ra \ell_j \gamma)$ be below their 
experimental upper 
bounds, one typically assumes that $\tilde{m}_L^2$ and all other 
slepton mass and trilinear coupling matrices are 
diagonal at the scale $M_X$. 
Such a situation occurs for instance in the CMSSM. 
Off-diagonal terms get induced at low energy scales radiatively, which 
explains their smallness. In this case a very good 
approximation for the typical SUSY mass appearing in 
eq.~(\ref{eq:BR}) is \cite{PPTS} 
%\be \label{eq:mS8} 
$m_S^8 = 0.5 \, m_0^2 \, m_{1/2}^2 \, 
(m_0^2 + 0.6 \, m_{1/2}^2)^2$,
%\ee
where $m_0$ is the universal scalar mass and $m_{1/2}$ is the universal 
gaugino mass at $M_X$. The well-known result for the slepton 
mass matrix entries is \cite{LFV} 
\be \label{eq:LFVI}
\hspace{-.5cm}
\left(\tilde{m}_L^2 \right)_{ij} 
= - \frac{(3 m_0^2 + A_0^2)}
{8 \, \pi^2 \, v_u^2} \, 
\left( m_D \, L \, m_D^\dagger \right)_{ij}\,,~\mbox{ where } ~\;
L_{ij} = \delta_{ij} \, \ln \frac{M_X}{M_i} \,.
\ee 
Here  $v_u = v \, \sin \beta$ and 
$A_0$ is the universal trilinear coupling. 
The logarithmic factor in eq.~(\ref{eq:LFVI}) 
takes into account the effect of running from the high scale $M_X$ 
to the scale of the respective heavy neutrino masses. 
We mentioned above the vertex between Higgs, leptons and the 
heavy neutrinos, which is the main aspect of see-saw. Its presence 
can be interpreted here in the form of a diagram with a 
slepton $j$ going into heavy (s)neutrino and Higgs(ino), 
which recombine into a slepton $i$. 
%This transition is necessary in the 
%$\ell_j \ra \ell_i \gamma$ process. 

Inserting the Casas-Ibarra parameterization from eq.~(\ref{eq:CI}) 
in $m_D \, m_D^\dagger$ reveals that, in general, 
in addition to the high energy parameters, LFV depends 
on all the parameters in the light neutrino mass 
matrix, including the 
Majorana phases, all three light neutrino masses 
and the mass ordering.

We stress here that due to the factorization of 
$\left(\tilde{m}_L^2 \right)_{ij} $ 
in a flavor and a SUSY term the 
ratios of the branching ratios are 
independent on the SUSY parameters. Hence they contain information on 
the flavor structure. For instance, 
\be \label{eq:doubleratio}
\frac{{\rm BR}(\mu \ra e \gamma)}{{\rm BR}(\tau \ra e \gamma)} \simeq 
\frac{1}{{\rm BR}(\tau \ra e \, \nu \overline{\nu})}
\left| 
\frac{\left(m_D \, L \, m_D^\dagger \right)_{12}}
{\left(m_D \, L \, m_D^\dagger \right)_{13}}
\right|^2~.
\ee
We will mostly consider these ratios of ratios from now on. Note 
that LFV (and later on leptogenesis) should be evaluated in 
the basis in which the heavy neutrino and the charged leptons 
are real and diagonal. If there are not 
diagonal, then $m_D$ should be 
replaced with $U_\ell^\dagger \, m_D \, V_R^\ast$, where 
$m_\ell \, m_\ell^\dagger = 
U_\ell \, (m_\ell^{\rm diag})^2 \, U_\ell^\dagger$ and 
$V_R^\dagger  \, M_R \, V_R^\ast$.

One simple example is the following: suppose both $m_D$ and $M_R$ 
obey a 2-3 exchange symmetry \cite{mutauseesaw}:  
\be \label{eq:mutauseesaw}
m_D = 
\left( 
\bad
a & b & b \\
d & e & f \\ 
d & f & e 
\ea
\right) \mbox{ and } 
M_R = 
\left( 
\bad
X & Y & Y \\
\cdot & Z & W \\
\cdot & \cdot & Z 
\ea
\right)\,.
\ee
Obviously $m_\nu$ will be $\mu$--$\tau$ symmetric, i.e., look 
like eq.~(\ref{eq:mnumutau}), in this case. Ignoring logarithmic 
corrections, one finds that 
$(m_D \, m_D^\dagger)_{21} = (m_D \, m_D^\dagger)_{31}$ and consequently 
BR$(\mu \ra e \gamma)/$BR$(\tau \ra e \gamma) \simeq 
1/{\rm BR}(\tau \ra e \, \nu \overline{\nu}) \simeq 5.7$. 
Up to the normalization factor the branching ratios are 
equal, which is so-to-speak a consequence of the fact 
that $\mu$--$\tau$ symmetry 
makes here no difference between muon and tau flavor.   

Another interesting case is connected with scaling and 
occurs when 
\be \label{eq:mdSSA}
m_D = 
\left(
\bad
a_1 & a_2 & a_3 \\
b & d & e \\
b/c & d/c & e/c 
\ea
\right)\,.
\ee
Interestingly, regardless of the form of $M_R$ the effective mass matrix 
obeys scaling with the scaling factor being the parameter $c$ in $m_D$, 
i.e., $m_\nu$ looks like eq.~(\ref{eq:SSA}). In this case, 
$(m_D \, L \, m_D^\dagger)_{21} /(m_D \, L \, m_D^\dagger)_{31}
= c^2$, which is nothing but $\cot^2 \theta_{23}$.

Recall the current limit of $1.2 \cdot 10^{-11}$ 
on BR$(\mu \ra e \gamma)$, and an expected improvement of two 
orders of magnitude on the limit of 
${\rm BR}(\tau \ra e \gamma) \le 1.1 \cdot 10^{-7}$. 
Therefore, in both examples it follows that 
$\tau \ra e \gamma$ will not be observed in a 
foreseeable future. The decay $\tau \ra \mu \gamma$ is 
not constrained.\\

Leaving this model-independent approach aside now, let us 
perform a GUT inspired estimate of the ratio of the branching 
ratios: 
suppose $m_D$ coincides 
with the mass matrix of up-type quarks $m_{\rm up}$.   
In addition, we will follow \cite{AFS} and assume that 
the mismatch between the left-handed rotations diagonalizing 
the Dirac-type 
neutrino mass matrix $m_D$ and the mass matrix of 
charged leptons $m_\ell$ is the same as the mismatch of 
the left-handed rotations diagonalizing 
the up-type and down-type quark matrices, i.e., is 
given by $V_{\rm CKM}$. This includes the special case 
in which $m_D = m_{\rm up}$ is diagonal and $m_\ell$ is diagonalized 
by the CKM matrix. This in turn occurs in a scenario leading to 
quark-lepton complementarity \cite{QLC,QLC0}, sometimes 
called QLC 1. In either realization of this possibility, 
heavy neutrino masses very similar 
to the ones in eq.~(\ref{eq:hie2}) will result. 
The overall result is that  
$m_D \, m_D^\dagger \simeq V_{\rm CKM}^\dagger  
\, {\rm diag}(m_u^2, m_c^2, m_t^2) \, V_{\rm CKM}$. 
We will adopt the Wolfenstein parameterization of the CKM matrix 
\cite{Wolf}:
\be
V_{\rm CKM} = 
\left(
\begin{array}{ccc}
1 - \lambda^2/2 & \lambda
& A \, \lambda^3 \, (\rho - i \eta)\\
-\lambda & 1 - \lambda^2/2 & A \, \lambda^2 \\
A \, \lambda^3 \, (1 - \rho + i\eta) & -A \, \lambda^2 & 1
\end{array}
\right)\,.
\label{eq:Wolf}
\ee
Here $A \simeq 0.82$, $\lambda \simeq 0.23$, 
$\rho \simeq 0.23$ and $\eta \simeq 0.35$ \cite{PDG}.  
Taking into account that the up-type quark masses satisfy 
$m_u : m_c : m_t \simeq \lambda^8 : \lambda^4 : 1$, we find  
\begin{eqnarray}
{\rm BR}(\mu \ra e \gamma) 
& \propto&  A^4 
\left( \eta^2 + (1 - \rho)^2 \right) \, \lambda^{10} \,, \\
{\rm BR}(\tau \ra e \gamma) 
& \propto & {\rm BR}(\tau \ra e \, \nu \overline{\nu}) \,  A^2 
\left( \eta^2 + (1 - \rho)^2 \right) \, \lambda^6 ~,\\
 {\rm BR}(\tau \ra \mu \gamma) 
& \propto & {\rm BR}(\tau \ra \mu \, \nu \overline{\nu}) \, 
A^2 \,  \lambda^{4} ~.
\end{eqnarray}
The relative size of the branching ratios can very well 
be described by  
\be \label{eq:Idom}
{\rm BR}(\mu \ra e \gamma) : {\rm BR}(\tau \ra e \gamma) : 
{\rm BR}(\tau \ra \mu \gamma) 
\simeq \lambda^5 : \lambda^2 : 1\,. \quad
\ee 
Here we have taken into account the normalization factors 
BR$(\tau \ra e \, \nu \overline{\nu}) \simeq {\rm BR}
(\tau \ra \mu \, \nu \overline{\nu}) \sim \lambda$. 
The relation in eq.~(\ref{eq:Idom}) 
implies that if ${\rm BR}(\mu \ra e \gamma)$ lies close to its current 
upper limit, then both $\tau \ra e \gamma$ and $\tau \ra \mu \gamma$ 
decays are observable. 
To give a feeling of the numerical 
values, we can use the parameters  
$m_0 = 100$ GeV, $m_{1/2} = 600$ GeV and $A_0 = 0$, for which 
${\rm BR}(\mu \ra e \gamma) \simeq 5 \cdot 10^{-19} 
\, \tan^2 \beta$.  

Again, we can consider the situation in realistic SUSY 
$SO(10)$ models. Recently a comparison of the predictions for LFV was 
performed in ref.~\cite{carl}. Table \ref{tab:so10_carl} summarizes 
the findings, where we have for convenience 
rewritten the numerical values from 
\cite{carl} in terms of powers of $\lambda$. 
Note that only in one model $\mu \ra e \gamma$ is not 
the rarest decay, and that the ratio of $\tau \ra e \gamma$ 
and $\tau \ra \mu \gamma$ is usually not too far away from 
our naive estimate in eq.~(\ref{eq:Idom}). 
In general the branching ratio for $\tau \ra \mu \gamma$ is 
the largest. The prediction for $\mu \ra e \gamma$ in the models 
CM (roughly $8 \cdot 10^{-19} \, \tan^2 \beta$ 
for $m_0 = 100$ GeV, $m_{1/2} = 600$ GeV and $A_0 = 0$)
and CY (roughly $2 \cdot 10^{-19} \, \tan^2 \beta$) 
is very close to our naive estimate. The other 
models predict a sizably larger branching ratio, BR$(\mu \ra e \gamma)$ 
for DR is more than two orders of magnitude larger, 
whereas model AB (GK) predict a branching ratio larger by five (six) 
orders of magnitude. 
%Hence, as shown in  
%ref.~\cite{carl}, if BR$(\mu \ra e \gamma)$ 
%is less than $10^{-13}$ the 
%models AB and GK in table \ref{tab:so10_carl} are eliminated. 

\begin{table}
\caption{Higgs content, predicted mass $M_1$ of the
lightest right-handed neutrino, BR$(\mu \ra e \gamma)$ divided by 
$\tan^2 \beta$ for $m_0 = 100$ GeV, $m_{1/2} = 600$ GeV, $A_0 = 0$, 
and the ratio of 
BR$(\mu \ra e \gamma) : {\rm BR}(\tau \ra e \gamma) : 
{\rm BR}(\tau \ra \mu \gamma)$ in various SUSY $SO(10)$ models. 
The prediction for $|U_{e3}|$ is also given. 
Taken from \protect\cite{carl} and slightly modified.}
\label{tab:so10_carl} 
\begin{tabular}{ccccccc} %\hline \hline 
 & AB \cite{AB}  & CM \cite{CM} & CY \cite{CY} 
& DR \cite{DR} & GK \cite{GK} & naive \\ \hline
Higgs 
& ${\bf 10}$, ${\bf 16}$, $\overline{\bf 16}$, ${\bf 45} $
& ${\bf 10}$, $\overline{\bf 126}$ 
& ${\bf 10}$, $\overline{\bf 126}$ 
& ${\bf 10}$, ${\bf 45}$ 
& ${\bf 10}$, ${\bf 120}$, $\overline{\bf 126}$ & ``${\bf 10}$'' \\ \hline
$M_1$ [GeV] & $4.5 \cdot 10^{8}$ & 
$1.1 \cdot 10^{7}$ & $2.4 \cdot 10^{12}$  & $1.1 \cdot 10^{10}$
& $6.7 \cdot 10^{12}$ & $2.0 \cdot 10^{5}$\\ \hline
$|U_{e3}|$  & 0.05 & 0.11 & 0.05 & 0.05 & 0.02 & -- \\ \hline
$\frac{\D {\rm BR}(\mu \ra e \gamma)}{\D \tan^2 \beta}$ 
& $5 \cdot 10^{-14}$ & $8 \cdot 10^{-19}$ 
& $2 \cdot 10^{-19}$ & $1 \cdot 10^{-16}$ & $2 \cdot 10^{-13}$ 
&  $5 \cdot 10^{-19}$ \\  \hline
ratio & $\lambda^2 : \lambda^3 : 1 $ 
& $\lambda^7 : \lambda^3 : 1 $ 
& $\lambda^4 : \lambda^3 : 1 $
& $\lambda^5 : \lambda^3 : 1 $
& $\lambda : \lambda : 1 $ 
& $\lambda^5 : \lambda^2 : 1 $
%BR$(\mu \ra e \gamma) : {\rm BR}(\tau \ra e \gamma)$ 
%& & & & & & \\
%$: {\rm BR}(\tau \ra \mu \gamma)$ & 
%{\raisebox{1.5ex}[-1.5ex]{$\lambda^2 : \lambda^3 : 1 $ }}
%& {\raisebox{1.5ex}[-1.5ex]{$\lambda^7 : \lambda^3 : 1 $}} 
%& {\raisebox{1.5ex}[-1.5ex]{$\lambda^4 : \lambda^3 : 1 $ }}
%& {\raisebox{1.5ex}[-1.5ex]{$\lambda^5 : \lambda^3 : 1 $ }}
%& {\raisebox{1.5ex}[-1.5ex]{$\lambda : \lambda : 1 $ }} 
%&  {\raisebox{1.5ex}[-1.5ex]{$\lambda^5 : \lambda^2 : 1 $}}
\end{tabular}
\end{table}

\subsection{\label{sec:etaB}Leptogenesis} 
See-saw is connected to heavy particles, and heavy masses correspond 
in cosmology to early times. The see-saw vertex 
of leptons, Higgs and heavy neutrinos shows up here in the form 
of a decay of the heavy neutrinos \cite{lepto}. 
The decay asymmetry is then (for a recent review, see \cite{D}) 
\bea 
\label{eq:epsIal}
\hspace{-2.28cm}\varepsilon_i^\alpha \D  
= \frac{\D \Gamma (N_i \ra \Phi \, \bar{l}_\alpha) -
\Gamma (N_i \ra \Phi^\dagger \, l_\alpha)}
{\D \Gamma (N_i \ra \Phi \, \bar{l}) + 
       \Gamma (N_i \ra \Phi^\dagger \, l)}  \\
\,=\, \D \frac{1}{8 \pi \, v_u^2} \, 
\frac{1}{(m_D^\dagger \, m_D)_{ii}}  
\, \sum\limits_{j \neq i} 
{\rm Im} \,\Big[ (m_D^\dagger)_{i \alpha} \, 
(m_D)_{\alpha j} \, 
\big(m_D^\dagger \, m_D \big)_{i j}\Big] \, 
f(M_j^2/M_i^2)~~~~~~~ \\ 
\D \,~~~+\, 
 \frac{1}{8 \pi \, v_u^2} \, 
\frac{1}{(m_D^\dagger \, m_D)_{ii}}  
\, \sum\limits_{j \neq i} 
{\rm Im} \Big[ (m_D^\dagger)_{i \alpha} \, 
(m_D)_{\alpha j} \, 
\big(m_D \, m_D^\dagger \big)_{i j} \Big]\, 
\frac{1}{1-M_j^2/M_i^2}
\,,
\eea
where    
\be
\D 
f(x) = 
\sqrt{x} \, \left( 
\frac{2}{1 - x} - \ln \left( \frac{1+x}{x}  \right) 
 \right) \,.
\ee
We have indicated here that flavor 
effects \cite{flavor_flav,flavor_others,Bari,petcov_flav,br,sasha} 
might play a role, i.e., $\varepsilon_i^\alpha$ describes 
the decay of the heavy neutrino 
of mass $M_i$ into leptons of flavor $\alpha = e, \mu, \tau$. 
In the case when the lowest-mass heavy 
neutrino is much lighter than the other two, i.e., 
$M_1 \ll M_{2,3}$, the lepton asymmetry is dominated by the decay 
of this lightest neutrino and $f(M_j^2/M_1^2) 
\simeq - 3 \, M_1/M_j$. In addition 
the last terms in eq.~(\ref{eq:epsIal}) are 
suppressed by an additional power of $M_1/M_j$. Note that the 
second term in eq.~(\ref{eq:epsI}) vanishes when summed over flavors 
$\alpha$:
\bea 
\label{eq:epsI}
\hspace{-1.28cm}\varepsilon_i \D  
= \sum\limits_\alpha 
\frac{\D \Gamma (N_i \ra \Phi \, \bar{l}_\alpha) -
\Gamma (N_i \ra \Phi^\dagger \, l_\alpha)}
{\D \Gamma (N_i \ra \Phi \, \bar{l}) + 
       \Gamma (N_i \ra \Phi^\dagger \, l)} 
\equiv \frac{\D \Gamma (N_i \ra \Phi \, \bar{l}) -
\Gamma (N_i \ra \Phi^\dagger \, l)}
{\D \Gamma (N_i \ra \Phi \, \bar{l}) + 
       \Gamma (N_i \ra \Phi^\dagger \, l)}  \\
\,=\, \D 
\frac{1}{8 \pi \, v_u^2} \, 
\frac{1}{(m_D^\dagger \, m_D)_{ii}}  
\, \sum\limits_{j \neq i} 
{\rm Im} \,
\Big[ \big(m_D^\dagger \, m_D \big)^2_{i j}\Big] 
\, f(M_j^2/M_i^2)
~.
\eea
The expressions we gave for the decay asymmetries are valid in case 
of the MSSM. Their flavor structure is however 
identical to the case of just the Standard Model. Also important 
in leptogenesis are the effective mass parameters responsible 
for the wash-out. Focussing on the case of the heavy neutrino 
$M_1$ being relevant for leptogenesis, every decay asymmetry 
$\varepsilon_1^\alpha$ is washed out by 
\be
\tilde{m}_1^\alpha = 
\frac{(m_D^\dagger)_{1 \alpha}\, (m_D)_{\alpha 1}}{M_1}~,
\ee
and the wash-out can be estimated by inserting this parameter 
in the function \cite{flavor_flav}
\be
\eta(x) \simeq 
\left( 
\frac{8.25 \cdot 10^{-3}~{\rm eV}}{x} + 
\frac{x}{2 \cdot 10^{-4}~{\rm eV}} 
\right)^{-1}~.
\ee
The final baryon asymmetry is 
\be
\hspace{-1.18cm}Y_B \simeq \left\{ 
\baz
-0.01 \, \varepsilon_1 \, \eta(\tilde{m}_1) 
& \mbox{one-flavor}~,\\[0.2cm]
-\frac{12}{37 \, g^\ast} 
\left( 
(\varepsilon_1^e + \varepsilon_1^\mu) \, \left(\frac{417}{589} 
(\tilde{m}_1^e + \tilde{m}_1^\mu) \right) + 
\varepsilon_1^\tau \, \left(\frac{390}{589} 
(\tilde{m}_1^\tau) \right) \right) 
& \mbox{two-flavor}~,\\[0.2cm]
-\frac{12}{37 \, g^\ast} 
\left( 
\varepsilon_1^e \, \left(\frac{151}{179} 
\tilde{m}_1^e  \right) + 
\varepsilon_1^\mu \, \left(\frac{344}{537} 
\tilde{m}_1^\mu \right) + 
\varepsilon_1^\tau \, \left(\frac{344}{537} 
(\tilde{m}_1^\tau) \right) \right) & \mbox{three-flavor}~.
\ea \right.
\ee 
Here $g^\ast = 228.75$ and we gave the expressions valid 
in the case of one-, two- and three-flavored 
leptogenesis. The three-flavor case occurs for 
$M_1 \, (1 + \tan^2 \beta) \le 10^9$ GeV, 
the one-flavor case for 
$M_1 \, (1 + \tan^2 \beta) \ge 10^{12}$ GeV, 
and the two-flavor case applies in between. The quantity 
$Y_B$ is defined as the number density of baryons divided by the 
entropy density: $Y_B = n_B/s$, which is related 
to $\eta_B = n_B/n_\gamma$ 
via $\eta_B = 7.04 \, Y_B$. The measured value is 
$Y_B = (0.87 \pm 0.03) \cdot 10^{-10}$ \cite{wmap}.  

Much activity has recently been spent on the 
implications of flavor effects 
\cite{flavor_flav,flavor_others,Bari,petcov_flav,br,sasha}. 
Neglecting flavor effects usually changes the predictions for $Y_B$ 
by an amount of order $10 \%$, but cases with 
discrepancies of several orders of magnitude are possible.  
The main issue of 
flavor effects, overlooked for many years, is that in the thermal 
plasma rates of processes like 
$q_L \, t_R \leftrightarrow \ell_{\alpha} \, \overline{\alpha_R}$ can 
be larger than the Hubble parameter. E.g., for $\alpha = \tau$ 
this happens in the SM for $T \le 10^{12}$ GeV. The process 
is thus ``in equilibrium'' 
and the tau flavor is distinguishable from the other flavors. We have 
to use now  $\varepsilon_1^\tau$ and $\varepsilon_1^{e + \mu}$ instead 
of $\varepsilon_1$. 
For the MSSM the Yukawa couplings $y$ for the process are 
replaced by $y \rightarrow y \, \sqrt{1 + \tan^2 \beta}$ and the 
temperature for which $H < \Gamma$ is consequently 
$T \, (1 + \tan^2 \beta)$.\\

\begin{figure}
\begin{center}
\epsfig{file=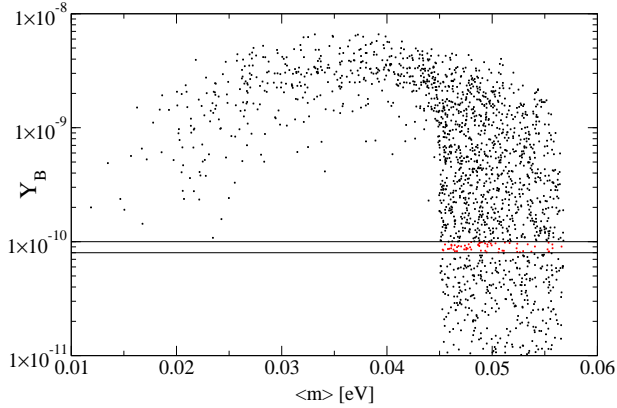,width=6.5cm,height=9cm,angle=270}
\caption{\label{fig:SSA}Correlation between the 
effective mass governing neutrinoless double beta decay and 
the baryon asymmetry. Taken from \protect\cite{SSA2}.}
\end{center}
\end{figure}
One interesting 
possible feature of leptogenesis 
is the connection of low energy CP violation to the 
CP violation necessary for leptogenesis. Without flavor 
effects $\varepsilon_1$ in eq.~(\ref{eq:epsI}) is relevant. 
After inserting 
the Casas-Ibarra parameterization in $\varepsilon_1$ 
it becomes clear that $U$, and therefore the low 
energy CP phases, do not show up in the decay asymmetry \cite{nocp,PPR}. 
Very frequently, however, specific models have a connection 
between high and low energy CP violation, originating 
from relations between mass matrix entries, zero textures, etc. 
There are countless examples for this, a recent one bases on scaling. 
The model from ref.~\cite{SSA2}, which bases on the flavor symmetry 
$D_4 \times Z_2$, results in diagonal charged lepton and 
heavy Majorana mass matrices, and 
\be
m_D = 
\left( 
\bad
a \, e^{i \phi} & b & 0 \\
0 & d & 0 \\
0 & e & 0  
\ea
\right)\,.
\ee
The effective mass matrix obeys scaling, with $c = d/e$, and  
due to the many zero textures there is only one CP phase. Recall 
that for scaling $m_3 = \theta_{13} = 0$, 
and therefore this phase is the Majorana 
phase in neutrinoless double beta decay and identical to the 
leptogenesis phase. Fig.~\ref{fig:SSA} shows the 
correlation between the effective mass and the baryon asymmetry.

In general, reproducing the observed value of $Y_B$, and its sign, 
is rarely a problem in models, 
including $SO(10)$ scenarios (see table \ref{tab:so10}). 
The naive GUT-inspired framework leading to the heavy 
neutrino masses in eq.~(\ref{eq:hie2}) and the ratio of 
branching ratios from eq.~(\ref{eq:Idom}) can also 
lead to leptogenesis \cite{AFS,QLC0}. 
However, recall that $M_1$ is typically well below $10^6$ GeV in 
eq.~(\ref{eq:hie2}). Therefore, it lies below the minimal 
mass value required for successful thermal leptogenesis, see below. 
Hence, tuning via CP phases is necessary in order 
to make $M_1$ and $M_2$ 
quasi-degenerate (see fig.~\ref{fig:qlc}) and to generate 
the baryon asymmetry via ``resonant leptogenesis'' \cite{resyb}.\\

\begin{figure}
\begin{center}
\epsfig{file=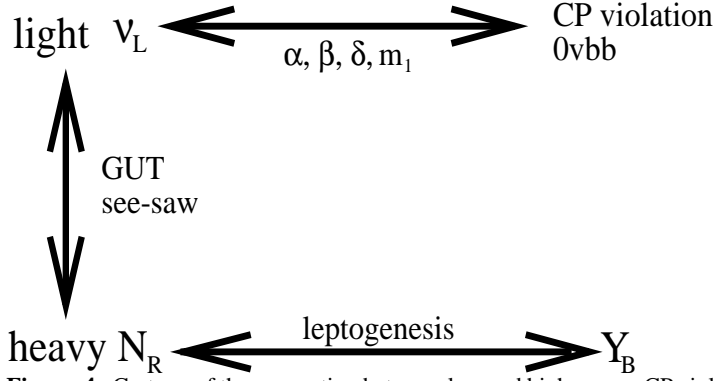,width=11cm,height=5cm}
\caption{\label{fig:pri}Cartoon of the connection between 
low and high energy 
CP violation. There is no direct link between low energy 
CP violation and the baryon 
asymmetry, a detour with model input/assumptions is required.}
\end{center}
\end{figure}

The general situation in what regards the connection of low and 
high energy CP violation slightly changes in case of 
flavored leptogenesis 
\cite{flavor_flav,flavor_others,Bari,petcov_flav,br}. 
This can be understood 
by inserting the Casas-Ibarra parametrization in the expression for 
the decay asymmetries $\varepsilon_1^\alpha$ in eq.~(\ref{eq:epsIal}). 
Note that they contain individual terms $(m_D)_{\alpha j}$ 
and $(m_D^\dagger)_{1\alpha}$. 
Consequently, terms in which $U$ explicitly shows up are present 
in $\varepsilon_1^\alpha$. 
Hence, if the low energy phases are non-trivial, they 
contribute to $Y_B$. 
Their effect can however be partly cancelled by the high energy 
CP phases in the complex orthogonal matrix $R$. 
In addition, flavored leptogenesis works 
perfectly well when the low energy phases 
vanish ($\alpha = \beta = \delta = 0$) \cite{sasha}. 
Connecting low and high energy CP violation is therefore similar, but 
not identical, to the case of unflavored leptogenesis: 
a certain amount of input/assumptions 
is necessary, see fig.~\ref{fig:pri}.\\ 

\begin{table}
\caption{\label{tab:fla}Comparison of the interplay of low and 
high energy 
neutrino physics for flavored and unflavored leptogenesis. Given are 
the upper limit on the smallest light neutrino mass, the lower limit 
on the smallest heavy neutrino mass, and if there is connection between 
high and low energy CP violation.}
\begin{tabular}{cccc} %\hline 
& mass of $m_1$ & mass of $M_1$ & low energy CP violation \\ \hline %\hline
No Flavor & $\ls 0.1$ eV & $\gs 10^9$ GeV &  no \\ %\hline
Flavor & free & $\gs 10^9$ GeV & maybe \\ %\hline  
\end{tabular}
\end{table}
The other interesting question in the framework of leptogenesis 
regards the required values of light and heavy neutrino masses. 
Most of the results 
depend on the wash-out and the Boltzmann-equations, and we refer 
to, e.g.,  
refs.~\cite{mMold,flavor_flav,flavor_others} for details. 
An important point is that there is 
an upper limit on $|\varepsilon_1|$ which decreases with 
the light neutrino mass scale \cite{DI}, a property not shared by 
$|\varepsilon_1^\alpha|$. Hence, there is an upper limit on neutrino 
masses for unflavored leptogenesis, but not for flavored leptogenesis. 
The upper limit on $M_1$ is basically not 
affected by the presence of flavor effects. 
Table \ref{tab:fla} summarizes the 
interplay of low and high energy neutrino physics in flavored and 
unflavored 
leptogenesis.

\begin{figure}
\begin{center}
\epsfig{file=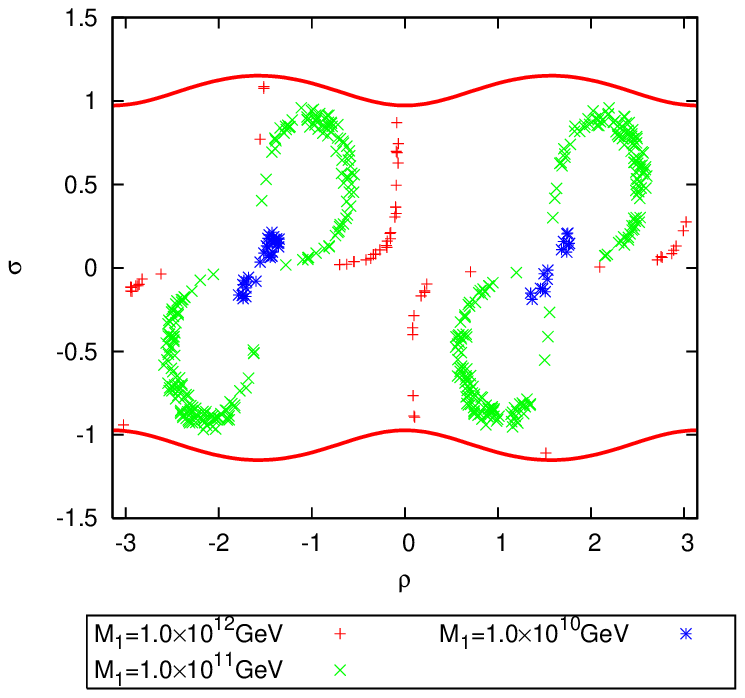,width=6cm,height=4cm}

\epsfig{file=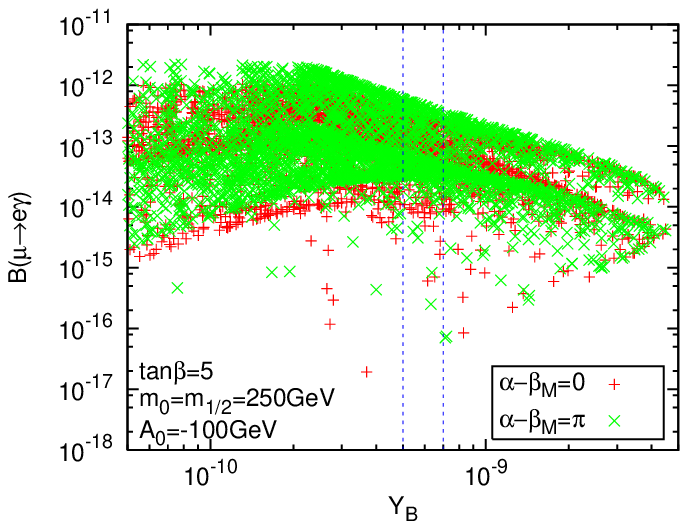,width=6cm,height=4cm} 
\epsfig{file=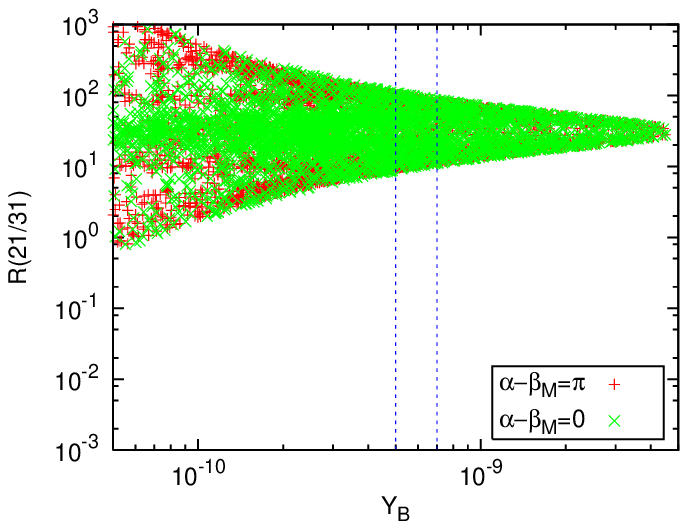,width=6cm,height=4cm}
\caption{\label{fig:PRST}Phenomenology of the scenario 
defined by eq.~(\ref{eq:Rsi}). Shown are the values for 
$\omega = \rho + i \sigma$ leading to successful (unflavored) 
leptogenesis, the correlation between $Y_B$ and the rate 
of $\mu \ra e \gamma$ and between $Y_B$ and 
BR$(\mu \ra e \gamma)$/BR$(\tau \ra e \gamma)$. 
%See \protect\cite{PRST} for details.
}
\end{center}
\end{figure}

\subsection{\label{sec:}Combining LFV and leptogenesis} 
One can try to combine now everything and try to understand 
the interplay of neutrino mass and mixing, LFV and leptogenesis 
\cite{davidson,ellis,PPR,comb,PRST,zero}. 
The following example \cite{PRST} shows that 
indeed interesting information on the flavor 
structure at high energy can be obtained and that 
the see-saw degeneracy can partly be broken: let us assume the 
SUSY parameters $m_0 = m_{1/2} = 250$ GeV and $A_0 = -100$ GeV. They 
correspond to\footnote{The LSP in this case is a neutralino 
of $\simeq 100$ GeV, the NLSP are a chargino and another neutralino 
with $\simeq 200$ GeV, squarks have masses in the range $400 \div 600$ 
GeV.} 
\be
{\rm BR}(\mu \ra e \gamma) \simeq 9.1 \cdot 10^{-9} 
\left|(m_D \, L \, m_D^\dagger)_{12}\right|^2 \frac{1}{v_u^4}
\tan^2 \beta\,.
\ee
Using the Casas-Ibarra parametrization implies that we can express 
$(m_D \, L \, m_D^\dagger)_{12}$ in terms of the heavy 
neutrino masses, the light neutrino parameters and the complex 
angles contained in $R$. 
The term proportional to $M_3$ will be the 
leading one. It can be found by setting 
$M_1 = M_2 = m_1 = 0$ and, for simplicity, inserting tri-bimaximal 
mixing:  
\bea \D 
(m_D \, L \, m_D^\dagger)_{12} \simeq 
-\frac 16 \, L_3 \, M_3 \, \sqrt{m_2} \, \cos \omega_{13} \, 
\cos \omega_{13}^\ast \\ \D 
\left(
\sqrt{6} \, e^{i(\alpha - \beta)} \, \sqrt{m_3} \, \cos \omega_{23} 
+ 2 \, \sqrt{m_2} \, \sin \omega_{23} 
\right) \, \sin \omega_{23}^\ast ~.
\eea
We have parameterized $R$ here as $R = R_{23} \, R_{13} \, R_{12}$. 
For a natural value of $M_3 = 10^{15}$ GeV it turns out that 
the branching ratio of $\mu \ra e \gamma$ is too large by at least 
three orders of magnitude. 
We can get rid of the potentially dangerous 
terms proportional to $M_3$ by setting $\omega_{13} = \pi/2$. 
If we would set $\omega_{23} = 0$ then  
terms of order $|U_{e3}| \, m_3 \, L_3 \, M_3 \, \cos \omega_{13} \, 
\cos \omega_{13}^\ast$ can lead to dangerously large 
${\rm BR}(\mu \ra e \gamma)$. For the value of $\omega_{13} = \pi/2$ 
the matrix $R$ simplifies to 
\be \label{eq:Rsi}
R = 
\left( 
\bad
0 & 0 & 1 \\
-\sin \omega & \cos \omega & 0 \\
-\cos \omega & -\sin \omega & 0 
\ea
\right)\mbox{ with } \omega = \omega_{12} + \omega_{23}~.
\ee
There is only one free complex parameter, 
which can be written as $\omega = \rho + i \sigma$ with real 
$\rho$ and $\sigma$. 
One can go on to study in this framework the constraints 
on $\omega$ from leptogenesis and also the implications for LFV, 
see fig.~\ref{fig:PRST}.

\section{\label{sec:other}Other See-Saws}
Up to now we have discussed the conventional, or type I see-saw, 
in which heavy neutrinos ($SU(2)_L$ singlets) are crucial. 
One special case in this framework is when $M_R$ is singular. 
In this ``singular see-saw'' one typically obtains light 
sterile neutrinos \cite{sing}. 

Apart from heavy neutrinos there are however 
other ways to generate the light neutrino mass 
matrix in eq.~(\ref{eq:mnu}). Instead of heavy $SU(2)_L$ singlets 
one could introduce heavy fermion triplets to the 
theory, which is called type III see-saw \cite{III}. 
More often studied is the case in which in addition to 
the heavy neutrinos one adds another singlet. In the 
$(\nu \, N^c \, S)$ basis one will obtain a general 
mass matrix of the form
\be \label{eq:cascade}
{\cal M} =  
\left(
\bad 
0 & m_D & 0 \\
m_D^T & 0 & m_{DS}^T \\
0 & m_{DS} & M_S
\ea
\right)~.
\ee
If $M_S \gg m_{DS} \gg m_D$ this is (for obvious reasons) 
called cascade, or sometimes double or inverse, 
see-saw \cite{cascade}, for which 
\be
m_\nu = m_D \, (m_{DS})^{-1} \, M_S \, (m_{DS}^T)^{-1} \, m_D~.
\ee
If one can realize that $m_D \propto m_{DS}$ then 
it follows that $m_\nu \propto M_S$ (``screening'') 
and one can blame the 
peculiar neutrino mixing structure entirely on the 
singlet sector \cite{screening}. 
Another possibility is to have in eq.~(\ref{eq:cascade}) an entry 
$\epsilon \, m_{DS}^T$ in the 13 element of ${\cal M}$. A contribution 
to the low energy mass matrix given by  
$-\epsilon \, (m_D + m_D^T)$ is the result \cite{barr}. 

\subsection{\label{sec:II}The Triplet or Type II See-Saw}
The most often studied variant of the see-saw\footnote{It is of 
course thinkable that all, or several, see-saw variants 
are simultaneously at work, or that something 
entirely different causes neutrino masses \cite{seesaw?}.} 
is the triplet, or type II, see-saw. 
A $SU(2)_L$ Higgs triplet $\Delta$ is introduced, which acquires 
a vev $v_L = \mu \, v_u^2 /M_{\Delta}^2$. Here $\mu$ is the 
doublet-doublet-triplet coupling parameter in the Higgs potential 
and $M_{\Delta}$ is the 
mass of the triplet, located around the same scale which heavy 
neutrinos have in the type I see-saw. The neutrino mass matrix 
is \cite{II}
\be \label{eq:mnuII}
m_\nu^{II} = v_L \, f_L~,
\ee
where $f_L$ is a Yukawa coupling matrix. Leptogenesis in the SM requires 
more than one triplet \cite{lepto_II}. In what regards LFV, 
one finds \cite{Rossi0}
\be\label{eq:LFVII}
\left(\tilde{m}_L^2 \right)_{ij}^{II} =  - 3 \, 
\frac{(3 m_0^2 + A_0^2)}{8 \, \pi^2 \, v_L^2} \, 
\left( m_\nu \, m_\nu^\dagger \right)_{ij} \, 
\ln \frac{M_X}{M_\Delta}\,.
\ee
The dependence of LFV on $m_\nu \, m_\nu^\dagger$ if a triplet is present 
has also been noticed in refs.~\cite{IILFV}\footnote{Regardless of the 
presence of a triplet, there is a 
contribution to LFV by massive neutrinos alone, which depends on 
$m_\nu \, m_\nu^\dagger$ as well. However, as well-known, this 
contribution is highly suppressed by a factor $(m_\nu/M_W)^4$.}. 
There is therefore a straight one-to-one correspondence between 
LFV and the directly measurable flavor structure of the 
neutrino mass matrix, so-to-speak ``minimal lepton flavor violation''.  
One can insert the parameterization of $U$ from 
eq.~(\ref{eq:Upara}) into $m_\nu = U \, m_\nu^{\rm diag} \, 
U^T$ and analyze the properties of the 
Hermitian matrix $h = m_\nu \, m_\nu^\dagger$ as a function 
of the known and unknown neutrino parameters. 
We note the following obvious but interesting differences 
\cite{Rossi0,Rossi1} with respect to the 
case of type I see-saw dominance, where, as we mentioned before 
(see the remarks above eq.~(\ref{eq:doubleratio})), 
LFV in general depends on all the parameters of $m_\nu$:    
\begin{itemize}
\item the Majorana phases drop out of $h$ 
and therefore do not influence LFV; 
\item the off-diagonal entries of $h$ do not depend on the overall 
neutrino mass scale, but only on $\dms$ and $\dma$. 
However, as can be seen from eq.~(\ref{eq:LFVII}), 
the overall neutrino mass scale appears in the 
branching ratios BR$(\ell_i \ra \ell_j \gamma)$ (though not in their 
ratios) via $v_L$; 
\item when varied over the the CP phase $\delta$,  
the moduli of the off-diagonal entries of $h$ are basically 
independent on the neutrino mass ordering. 
Their relative differences for normal and 
inverted mass orderings are of order $r \equiv \dms/\dma$ and therefore 
negligible. However, for fixed $\delta$ there can be differences: 
for instance, the result for $(m_\nu \, m_\nu^\dagger)_{12}$ 
in case of a normal (inverted) ordering and 
$\delta = 0$ is identical to the result 
for $(m_\nu \, m_\nu^\dagger)_{12}$ in case of an inverted (normal) 
ordering and $\delta = \pi$. 
\end{itemize} 
Measuring the branching ratios of the LFV decays 
$\ell_i \ra \ell_j \gamma$ 
will therefore teach us nothing about the neutrino 
properties that we could not learn from oscillation experiments. 
This is, of course, a consequence 
of the fact that both depend on the same quantity, namely  
$m_\nu \, m_\nu^\dag$. On 
the other hand, the neutrino parameters that are most difficult  
to determine -- the Majorana phases -- do not induce uncertainty 
in the predictions of the branching ratios. In addition, 
the ratio of branching ratios does not depend on the neutrino mass scale.

An immediate question one may ask is if the off-diagonal entries 
of $h = m_\nu \, m_\nu^\dagger$ (and therefore 
the branching ratios for 
$\ell_i \ra \ell_j  \gamma$ decays) can vanish \cite{Rossi1}. 
The analysis shows that  
\begin{itemize}
\item the quantity $h_{12}$ and therefore BR$(\mu \ra e \gamma)$ 
can vanish. Recall that the invariant describing 
CP violation in neutrino 
oscillations is $J_{\rm CP} \propto {\rm Im} \left\{ h_{12} \, 
h_{23} \, h_{31} 
\right\}$ (see eq.~(\ref{eq:jcp0})). Therefore, 
vanishing BR$(\mu \ra e \gamma)$ means the absence of 
CP violation in the case of type II dominance. 
The converse is, of course, not true. Note however that 
it is not possible to show experimentally that the 
branching ratio vanishes, and that 2-loop 
effects\footnote{In case of type I dominance the 
requirement of vanishing $(m_D \, m_D^\dagger)_{12}$ can 
lead via 2-loop effects to a lower limit 
on BR$(\mu \ra e \gamma)$, connected to the product of the 
branching ratios of $\tau \ra \mu \gamma$ and 
$\tau \ra e \gamma$ \cite{zero}.}  will 
induce small LFV even of $h_{12} = 0$. 
There is also a correlation between the 
neutrino mixing parameters which is easily obtained 
from $h_{12} = 0$ \cite{HR,Rossi1,h120}: 
\be \label{eq:12zero}
|U_{e3}| = \frac{1}{2} \, \frac{r \, \sin 2 \theta_{12} \, 
\cot \theta_{23} }{1 \mp r \, \sin^2 \theta_{12}} 
\simeq \frac{1}{2} \, r \, \sin 2 \theta_{12} \, 
\cot \theta_{23} =  0.016^{+0.013}_{-0.008}\, ,
\ee
where the $-$ sign is for the normal and the $+$ for the 
inverted neutrino mass ordering. Here $r = \dms/\dma$ is 
the (positive) ratio of the mass-squared differences; 
\item the quantity $h_{13}$ and therefore BR$(\tau \ra e \gamma)$ 
can vanish as well. Again, from eq.~(\ref{eq:jcp0}) we see that 
CP is conserved if $h_{13}=0$, and one can also obtain 
\be \label{eq:13zero}
|U_{e3}| = \frac{1}{2} \, \frac{r \, \sin 2 \theta_{12} \, 
\tan \theta_{23} }{1 \mp r \, \sin^2 \theta_{12}} 
\simeq \frac{1}{2} \, r \, \sin 2 \theta_{12}   
\tan \theta_{23} = 0.013^{+0.014}_{-0.006}~. 
\ee
From the above two formulae it 
is clear that BR$(\mu \ra e \gamma)$ and BR$(\tau \ra e \gamma)$ 
can vanish simultaneously only if $\theta_{23} = \pi/4$;  
\item the quantity $h_{23}$, and therefore 
BR$(\tau \ra \mu \gamma)$, cannot vanish. 
The reason is that $h_{23}=0$ would 
imply $\dms/\dma \simeq 1/\cos^2 \theta_{12} + {\cal O}(\theta_{13})$, 
in contradiction with experiment. 
\end{itemize}

In general, $h_{23}$ depends very little 
on $|U_{e3}|$ (the leading term with $|U_{e3}|$ is multiplied with 
small $r = \dms/\dma$) and is much larger than $h_{12}$ and $h_{13}$. 
While the leading term in $h_{23}$ is of order 
\dma, $h_{12}$ and $h_{13}$ are 
to leading order given by $\dma \, |U_{e3}|$ or 
\dms, depending on the magnitude of $|U_{e3}|$.  
If we adopt for simplicity tri-bimaximal mixing, we find 
\be
h_{12} = -h_{13} = 
\frac 13 \, \dms\,,\quad
h_{23} = \frac 16 \left(\pm 3 \, \dma - 2 \, \dms \right)\,,
\ee
where the plus (minus) sign is for the normal (inverted) neutrino 
mass ordering. 

\begin{figure}[t]
\begin{center}
\epsfig{file=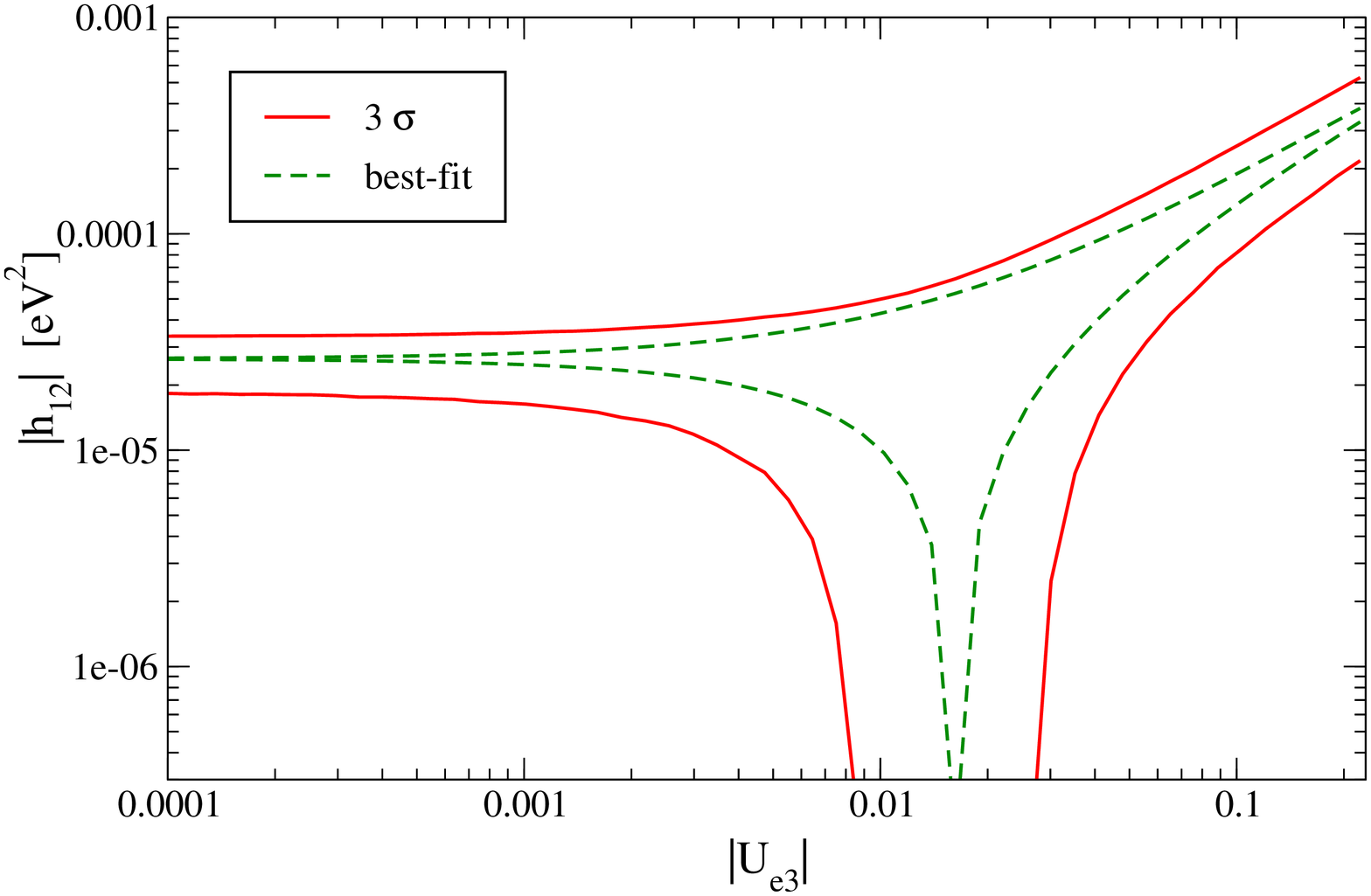,width=6.2cm,height=6cm}
\epsfig{file=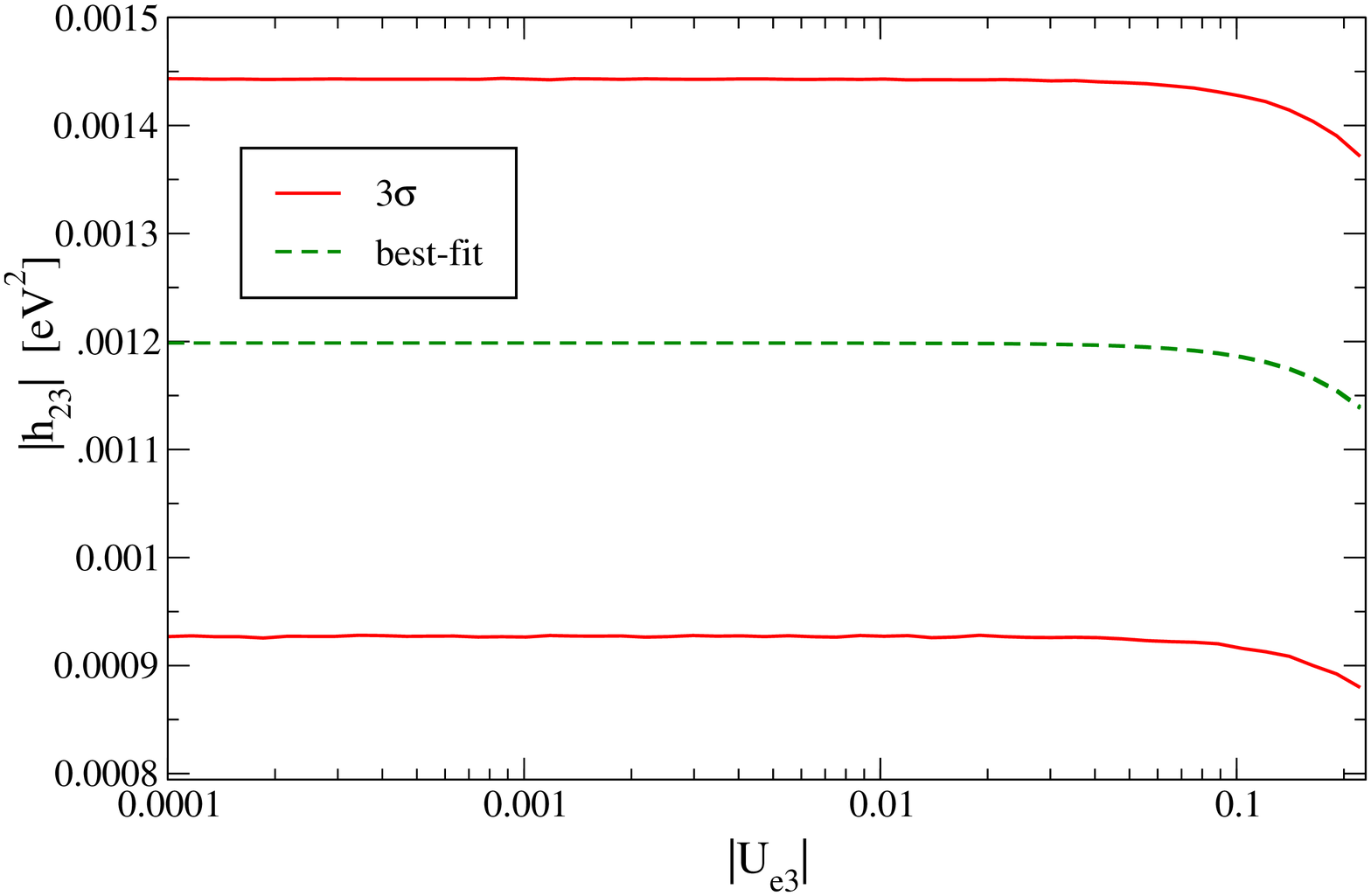,width=6.2cm,height=6cm}
\caption{\label{fig:h12h23}The values of $|h_{12}|$ (left panel)  
and $|h_{23}|$ (right panel, note the linear scale) 
as functions of $|U_{e3}|$, for all the other 
oscillation parameters varied within their 
allowed 3$\sigma$ ranges (solid curves) and 
for all the oscillation parameters 
except $\delta$ fixed to their best-fit values and only 
$\delta$ varied (dashed curves). }
\end{center}
\end{figure}

We show in fig.~\ref{fig:h12h23} the absolute values of 
$h_{12}$ and $h_{23}$ as functions of $|U_{e3}|$ (i) 
for all the other oscillation parameters varied 
within their allowed 3$\sigma$ ranges and (ii) for 
these parameters fixed to their best-fit values and 
only $\delta$ varied (see also 
ref.~\cite{Rossi1}). The element $|h_{13}|$ 
looks very much like $|h_{12}|$, which is due to the 
approximate $\mu$--$\tau$ 
symmetry implied by the neutrino data. We assumed the 
normal mass ordering in this figure, but, as mentioned above, 
the difference with regard to the 
case of the inverted ordering is negligible if the phase $\delta$ 
is varied.  

\begin{figure}[t]
\begin{center}
\epsfig{file=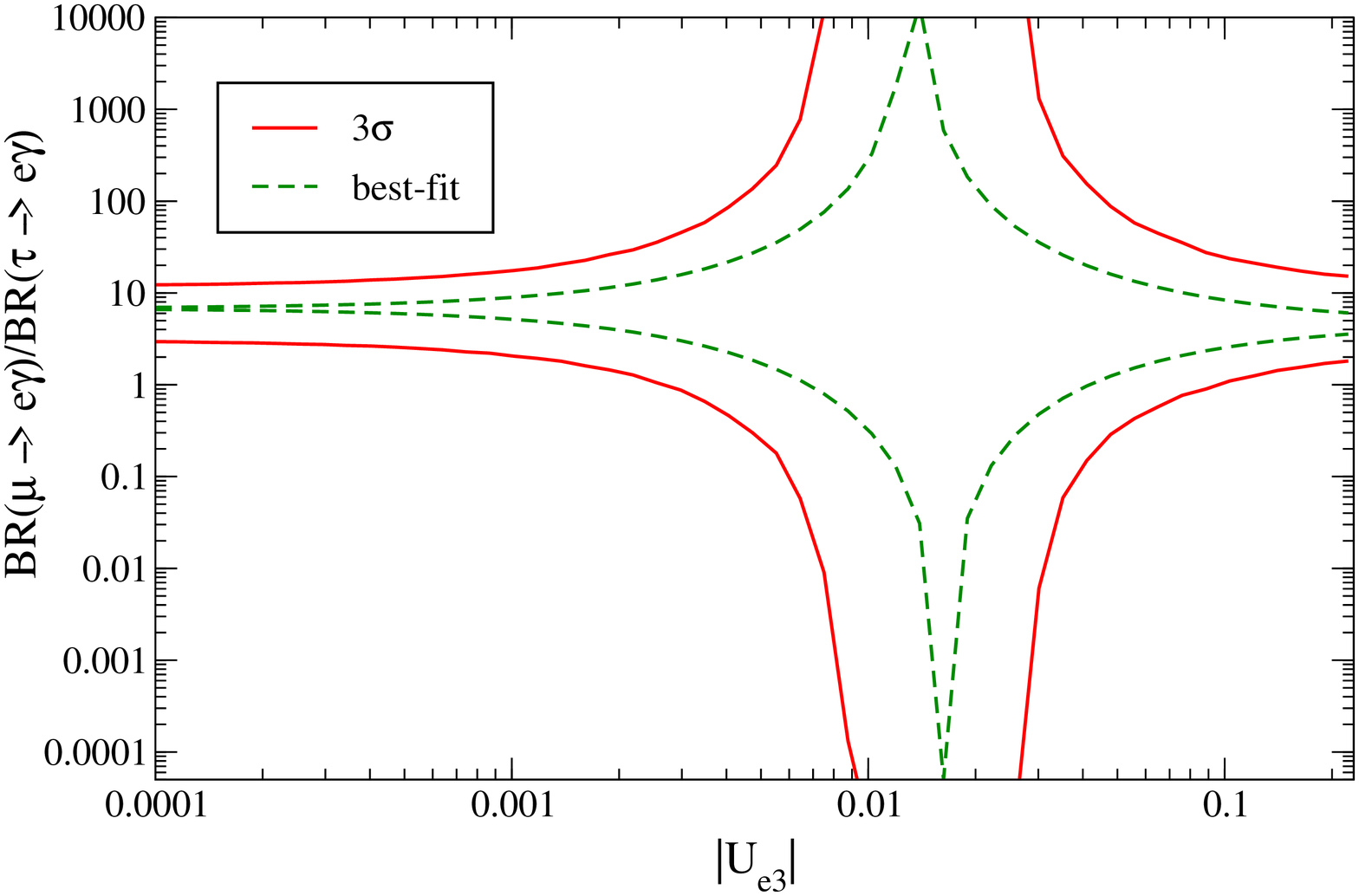,width=6.2cm,height=6cm}
\epsfig{file=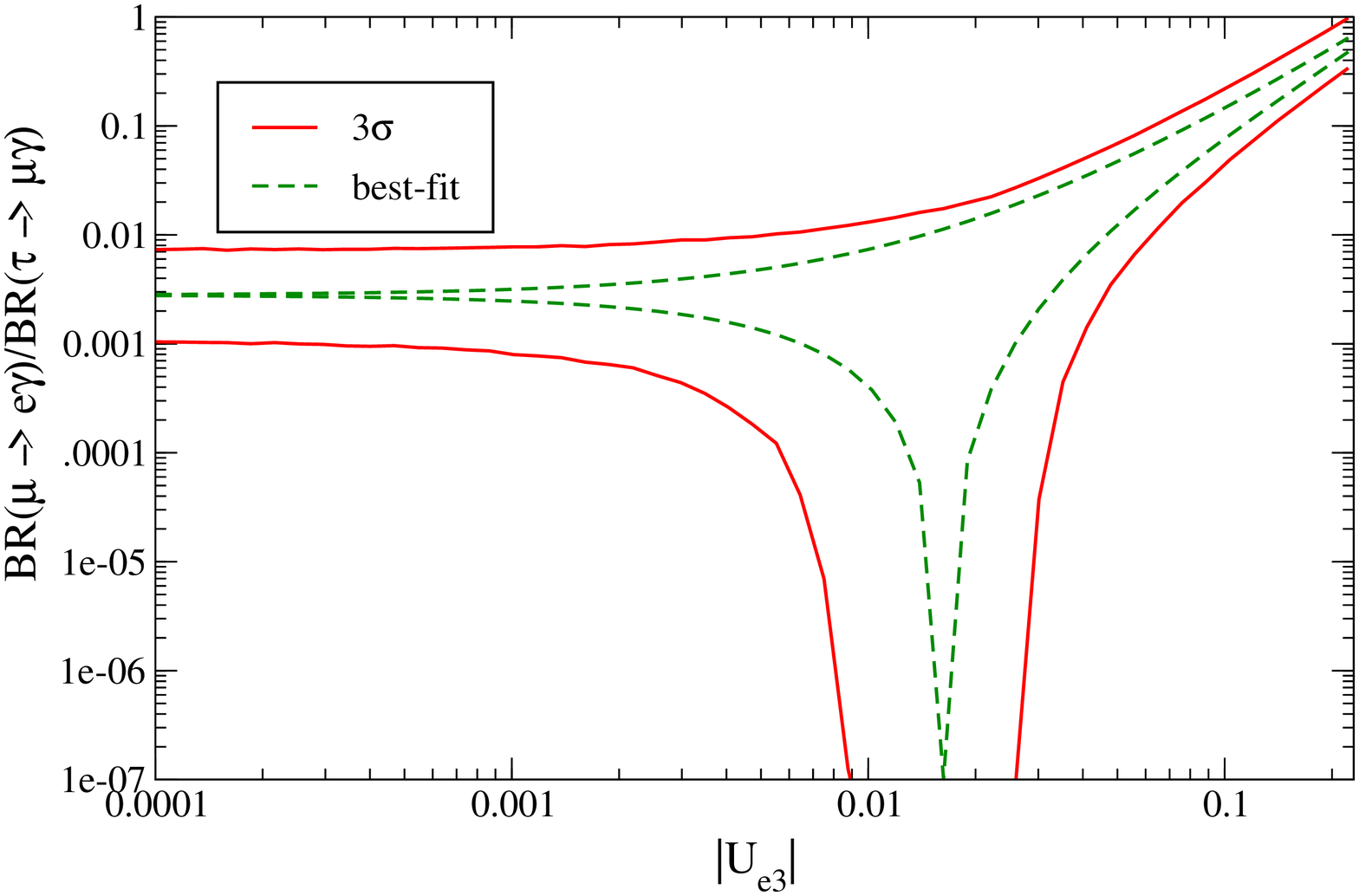,width=6.2cm,height=6cm}
\caption{\label{fig:hijratio}Type II dominance: the ratios 
BR$(\mu \ra e \gamma)/$BR$(\tau \ra e \gamma)$ and  
BR$(\mu \ra e \gamma)/$BR$(\tau \ra \mu \gamma)$ 
as functions of $|U_{e3}|$, for all the other 
oscillation parameters varied 
within their allowed 3$\sigma$ ranges (solid curves) 
and for all the 
oscillation parameters except $\delta$ fixed to their 
best-fit values and only $\delta$ varied (dashed curves). }
\end{center}
\end{figure}

Perhaps more interesting are the {\it ratios} of $|h_{ij}|^2$, which 
are directly proportional to the ratios of the branching ratios 
under discussion. If both $h_{12}$ and $h_{13}$ are 
not too small (i.e.~barring 
exact or almost exact cancellations between various 
terms contributing to 
these quantities), one finds, setting $|U_{e3}|$ to zero, 
\be \label{eq:IIR1213}
\frac{|h_{12}|^2}{|h_{13}|^2} = \frac{{\rm BR}(\mu \ra e  \gamma )}
{{\rm BR}(\tau \ra e  \gamma )}\,
{\rm BR}(\tau \ra e \, \nu  
\overline{\nu}) \, 
\simeq \cot^2 \theta_{23}\, ,
\ee
which is very close to one \cite{Rossi0}. 
The result for the ratio of ratios is 
the same as for the example based on scaling, see 
eq.~(\ref{eq:mdSSA}). Of course, if 
$h_{12}$ or $h_{13}$ becomes very 
small, this ratio can be arbitrarily large or small. 
For the ratio of $|h_{12}|^2$ and $|h_{23}|^2$ we get
\bea \D \label{eq:IIR1223}
\hspace{-.4cm}\frac{|h_{12}|^2}{|h_{23}|^2} = 
\frac{{\rm BR}(\mu \ra e  \gamma )}
{{\rm BR}(\tau \ra \mu  \gamma )} \,
{\rm BR}(\tau \ra \mu \, \nu  \overline{\nu}) \, 
\qquad\qquad\qquad
\qquad\qquad\qquad\quad
\vspace*{0.2cm}\\[0.2cm]
\D \,\simeq 
\frac{1}{\cos^2 \theta_{23}} \, |U_{e3}|^2 + 
\frac{\sin^2 2 \theta_{12}}{4 \, \sin^2 \theta_{23}} \, r^2 + 
2 \, \cos \delta \, \frac{\sin 2 \theta_{12}}
{\sin 2 \theta_{23}} \, 
r \, |U_{e3}|\,,
\eea 
which is rather small. 
Note that the maximal allowed value of $|U_{e3}|^2$ is roughly 
$\lambda^2$, while the value of 
$r^2 = (\dms/\dma)^2$ is approximately $\lambda^5$. 
Hence, for small $|U_{e3}|$ this ratio is 
given by $\lambda^5$, while for large $|U_{e3}|$ it is 
given by roughly $\lambda^2$. 
 We show in fig.~\ref{fig:hijratio} the two ratios 
$(|h_{12}|/|h_{13}|)^2 /0.178$ and $(|h_{12}|/|h_{23}|)^2 /0.174$ as 
functions of $|U_{e3}|$. These ratios are equal to 
BR$(\mu \ra e \gamma)/$BR$(\tau \ra e \gamma)$ and 
BR$(\mu \ra e \gamma)/$BR$(\tau \ra \mu \gamma)$, respectively. 
As in fig.~\ref{fig:h12h23}, we 
have either varied 
all the relevant parameters ($\theta_{12}$, 
$\theta_{23}$, \dms, \dma\,    
and $\delta$) within their allowed 3$\sigma$ ranges or 
fixed all these 
parameters except $\delta$ to their best-fit values, 
while allowing $\delta$ to vary.

Regarding the ratio of branching ratios, from eqs.~(\ref{eq:IIR1213}) 
and (\ref{eq:IIR1223}) one finds that  
\be \label{eq:Ddom}
\hspace{-1.18cm}
{\rm BR}(\mu \ra e \gamma) : {\rm BR}(\tau \ra e \gamma) : 
{\rm BR}(\tau \ra \mu \gamma) \simeq 
\left\{
\baz 
 \lambda : \lambda^2 : 1 & \mbox{for large } |U_{e3}|\,, \\ 
 \lambda^4 : \lambda^5 : 1 & \mbox{for small } |U_{e3}| \,. 
\ea 
\right.
\ee
Again, the normalization factors  
BR$(\tau \ra e \,  \nu  \overline{\nu}) \simeq {\rm BR}
(\tau \ra \mu \,  \nu  \overline{\nu}) 
\sim \lambda$ have been taken into account. The 
relation between $\mu \ra e \gamma$ and 
$\tau \ra e \gamma$ is the same as for $\mu$--$\tau$ 
symmetry, see eq.~(\ref{eq:mutauseesaw}). The branching ratio 
for $\tau \ra \mu \gamma$ is the largest one. It was also 
the largest one in the $SO(10)$ models (summarized in ref.~\cite{carl} 
and table \ref{tab:so10_carl}). We conclude that this is 
a quite generic and robust prediction. Observation different from 
this would mean that a combination of the type I and triplet see-saw 
(or something entirely different) causes $m_\nu$ and/or LFV. 
We can compare the different properties of LFV in the cases when 
either the type I see-saw term or the triplet term dominates 
in both $m_\nu$ and $\tilde{m}_L^2$. This is shown in 
table \ref{tab:IvsII}. Comparing the type II relation for the 
ratios of branching ratios with the $SO(10)$ results from table 
\ref{tab:so10_carl} shows that only model AB could be mimicked by 
a pure type II scenario. The relation $\lambda^2 : \lambda^3 : 1$ 
of this model can be obtained in a type II scenario if 
$|h_{12}|^2 /|h_{23}|^2 \simeq \lambda^3$, which can be obtained 
for a not too small $|U_{e3}| \simeq 0.07$. Therefore, it may be 
possible to distinguish see-saw variants with observables related 
to observables.

Let us compare the absolute magnitudes of the branching 
ratios in the 
cases of type I and type II dominance. Assuming that 
the logarithmic factors in $(\tilde{m}_L^2)_{ij}$ are the same, 
and using our simple GUT-inspired scenario from above 
(see eq.~(\ref{eq:Idom})), 
we obtain   
\be
\frac{{\rm BR}(\mu \ra e \gamma)|_{\rm type\,I\,dom.}}
{{\rm BR}(\mu \ra e \gamma)|_{\rm type\,II\,dom.}} \simeq 
\frac{A^4 \, \lambda^{10}}{(\dms/v_L^2)^2} 
\simeq 25 \, \left(\frac{v_L}{\rm eV}\right)^4\,,
\label{eq:ratio}
\ee
where we have used tri-bimaximal mixing to evaluate 
$h_{12}$. For $v_L \sim \sqrt{\dma}$ (normal or inverted 
hierarchy) we would expect a ratio 
of the order of $10^{-4}$ to $10^{-5}$, i.e.~in that case type II dominance 
would result in much larger LFV branching 
ratios than type I dominance. To be precise, with 
$m_0 = 100$ GeV, $m_{1/2} = 600$ GeV, $A_0 = 0$ and with 
tri-bimaximal mixing in a normal hierarchy we find 
BR$(\mu \ra e \gamma) \simeq 7 \cdot 10^{-14} \, \tan^2 \beta$. 
If $v_L$ approaches the eV scale, the two cases 
lead to branching ratios of the same order of magnitude. 
As mentioned above (see the discussion at the end of sec.~3.1), 
for the $SO(10)$ models CM and CY from table \ref{tab:so10_carl} 
a very similar ratio will hold. The other models (DR, AB and GK) 
have BR($\mu \ra e \gamma$) larger by two, five and six 
orders of magnitude, respectively. Recall that all of them have 
dominance of the type I see-saw term. 
%Since neither 
%$v_L$ nor the effective SUSY mass $m_S$ on which the LFV branching 
%ratios depend sensitively are currently known, a measurement of 
%BR$(\mu \ra e \gamma)$ (or of any other LFV branching ratio) by itself 
%would not be sufficient to unravel which see-saw 
%type (if any) dominates.\\ 

\begin{table}
\caption{\label{tab:IvsII}Comparison of general features of LFV 
in the cases when one of the two terms in the 
see-saw formula eq.~(\ref{eq:II}) 
dominates in both $m_\nu$ and $\tilde{m}_L^2$. 
For type I dominance (middle column), the  
entries marked with ``\,$^a\,$'' refer to the general case, in which  
$m_D \, m_D^\dagger \propto U \, \sqrt{m_\nu^{\rm diag}}  
\, R \, M_R \, R^\dagger \, \sqrt{m_\nu^{\rm diag}} \, U^\dagger$. 
The expectation given at the bottom and marked with ``\,$^b$\,'' 
assumes the GUT inspired relation $m_D \, m_D^\dagger =  
V_{\rm CKM}^\dagger  \, {\rm diag}(m_u^2, m_c^2, m_t^2) \, 
V_{\rm CKM}$. The superscript $^{\rm c}$ indicates that,  
if varied over the CP phase $\delta$, the 
neutrino mass ordering plays no role. 
More realistic models are given 
in table \ref{tab:so10_carl}.}
\begin{tabular}{ccc} 
 & Type I & Type II \\ \hline
relevant quantity & $ (m_D \, m_D^\dagger)_{ij} $ 
& $\left(m_\nu \, m_\nu^\dagger\right)_{ij}$ 
\\ \hline
does \underline{not} depend on & -- $^a$ 
& $\ba \mbox{Majorana phases} \\
\mbox{\scriptsize (and mass ordering)$^{\rm c}$} \ea$ \\ \hline
guaranteed & -- $^a$ & $\ba 
{\rm BR}(\tau \ra \mu \gamma) \neq 0 
\ea $ \\ \hline 
$\ba \mbox{expectation for} \\ 
{\scriptsize 
{\rm BR}(\mu \ra e \gamma) : 
{\rm BR}(\tau \ra e \gamma) : {\rm BR}(\tau \ra \mu \gamma) }\ea $ 
& $ \lambda^5 : \lambda^2 : 1 $ $^b$ & 
$ \left\{
\baz 
 \lambda : \lambda^2 : 1 & \mbox{for large } |U_{e3}| \\ 
 \lambda^4 : \lambda^5 : 1 & \mbox{for small } |U_{e3}|  
\ea 
\right.
$
\end{tabular}
\end{table}

\subsection{Triplet See-Saw and Type I See-Saw}
One often studies the case in which both the triplet term and 
the conventional see-saw term are 
present: 
%\footnote{Sometimes in the literature the eq.~(\ref{eq:II}) 
%is also called type I + II see-saw formula.}: 
\be \label{eq:II}
m_\nu = v_L \, f_L - m_D \, M_R^{-1} \, m_D^T~.
\ee
Dominance of one of the terms in both $m_\nu$ and 
$\left(\tilde{m}_L^2 \right)_{ij} $ corresponds to the 
situations discussed above. 
Leptogenesis has been studied in this framework \cite{IIYB}. 
Very often a discrete left-right symmetry is assumed, for which 
$f_L \propto M_R$ holds. 
Anyway, the neutrino mass 
matrix is a sum of two terms now, which can be a reason for 
the peculiar mixing structure of the neutrinos \cite{LR}. 
For instance, recall the tri-bimaximal mass matrix in 
eq.~(\ref{eq:mnutbm}). In a normal hierarchy the term 
proportional to $m_1$ vanishes, and we are left with two simple 
matrices, which could stem from either $m_\nu^{II}$ or 
from the conventional term $m_D \, M_{R}^{-1} \, m_D^T$.    
LFV will be complicated by the fact that the slepton mass 
matrix obtains contributions from both terms. It is 
in this case a sum of eq.~(\ref{eq:LFVI}) and (\ref{eq:LFVII}), 
therefore interference can occur. If the triplet term was known, one 
could subtract it from $m_\nu$ to obtain 
\be
\label{eq:Xnu}
\hspace{-1.15cm}X_\nu \,\equiv\, m_\nu - v_L \, f_L = 
- m_D \, M_{R}^{-1} \, m_D^T \,,\quad
\mbox{diagonalized as}\, \quad
X_\nu \,=\, V_\nu^\ast \, X_\nu^{\rm diag} V_\nu^\dagger
\ee
with a unitary matrix $V_\nu$. Now, 
in analogy to the Casas-Ibarra parameterization, 
one can parameterize the Dirac mass matrix as \cite{AR}
\be 
m_D = i \, V_\nu^\ast \, \sqrt{X_\nu^{\rm diag}} \, R \, \sqrt{M_R}~.
\ee 
Here $R$ is again an arbitrary complex and orthogonal matrix, in 
analogy to the $R$ in the Casas-Ibarra parametrization. 
Simple examples how to study LFV, leptogenesis and neutrino 
mass and mixing in this framework are given in ref.~\cite{AR}.

\section{\label{sec:concl}Summary}
The neutrino mass matrix and its origin are an exciting field 
of research, with overlap to many fields of (astro)particle physics, 
including SUSY phenomenology and cosmology. The see-saw 
mechanism (or any one of its many variants) and its 
challenging reconstruction represent the 
crucial link between these fields. Future data will help us draw
a clearer picture of the flavor structure in the lepton sector, and 
if we are lucky we could even discriminate between different see-saw 
variants. 
The hope is that in the not too far future 
only a limited number of theories/scenarios 
survive which are able to explain all observations.

\acknowledgements
I would like to thank my co-authors for fruitful collaborations, 
Carl Albright for careful reading of the manuscript and 
I am grateful to Evgeny Akhmedov for noting some errors in 
previous versions of the draft. 
I thank the organizers of WHEPP-X for their hospitality.   
This work was supported by the Deutsche Forschungsgemeinschaft 
in the Transregio 27 ``Neutrinos and beyond -- weakly interacting 
particles in physics, astrophysics and cosmology''. 
%\newpage


\begin{thebibliography}{99}


\bibitem{APS}R.~N.~Mohapatra {\it et al.}, 
{\it Rept.\ Prog.\ Phys.} {\bf 70}, 1757 (2007)
  %%CITATION = HEP-PH 0510213;%%

R.~N.~Mohapatra and A.~Y.~Smirnov,
  %``Neutrino mass and new physics,''
   {\it Ann.\ Rev.\ Nucl.\ Part.\ Sci.} {\bf 56}, 569 (2006)  
  %%CITATION = HEP-PH 0603118;%%

A.~Strumia and F.~Vissani,
  %``Neutrino masses and mixings and,''
  hep-ph/0606054 
  %%CITATION = HEP-PH 0606054;%%

\bibitem{jcp}
G.~C.~Branco {\it et al.}, %R.~Gonzalez Felipe, F.~R.~Joaquim, I.~Masina, M.~N.~Rebelo and C.~A.~Savoy,
  %``Minimal scenarios for leptogenesis and CP violation,''
  {\it Phys.\ Rev.}  {\bf D67}, 073025 (2003) 
%  [arXiv:hep-ph/0211001].
  %%CITATION = PHRVA,D67,073025;%%

\bibitem{newdata}M.~C.~Gonzalez-Garcia and M.~Maltoni,
  %``Phenomenology with Massive Neutrinos,''
  arXiv:0704.1800 [hep-ph]
  %%CITATION = ARXIV:0704.1800;%%

\bibitem{tbm}
P.~F.~Harrison, D.~H.~Perkins and W.~G.~Scott,
  %``Tri-bimaximal mixing and the neutrino oscillation data,''
  {\it Phys.\ Lett.} {\bf B530}, 167 (2002) and {\bf B535}, 163 (2002)
  %%CITATION = HEP-PH 0202074;%%
%  {\it Phys.\ Lett.} {\bf B535}, 163 (2002)
  %%CITATION = HEP-PH 0203209;%%

Z.~Z.~Xing,
  %``Nearly tri-bimaximal neutrino mixing and CP violation,''
  {\it Phys.\ Lett.} {\bf B533}, 85 (2002)
  %%CITATION = HEP-PH 0204049;%%

  X.~G.~He and A.~Zee,
  %``Some simple mixing and mass matrices for neutrinos,''
  {\it Phys.\ Lett.} {\bf B560}, 87 (2003)
  %%CITATION = HEP-PH 0301092;%%

see also 
L.~Wolfenstein,
  {\it Phys.\ Rev.} {\bf D18}, 958 (1978)
  %%CITATION = PHRVA,D18,958;%%

Y.~Yamanaka, H.~Sugawara and S.~Pakvasa,
  %``Permutation Symmetries And The Fermion Mass Matrix,''
  {\it Phys.\ Rev.}  {\bf D25}, 1895 (1982)

%  [Errat.\  {\bf D29}, 2135 (1984)]
  %%CITATION = PHRVA,D25,1895;%%

\bibitem{PRW}S.~Pakvasa, W.~Rodejohann and T.~J.~Weiler,
  %``TriMinimal Parametrization of the Neutrino Mixing Matrix,''
  {\it Phys.\ Rev.\ Lett.} {\bf 100}, 111801 (2008) 
  %%CITATION = ARXIV:0711.0052;%%

\bibitem{CR} S.~Choubey and W.~Rodejohann,
  %``A flavor symmetry for quasi-degenerate neutrinos: L(mu)-L(tau),''
  {\it Eur.\ Phys.\ J.}  {\bf C 40}, 259 (2005)
%  [arXiv:hep-ph/0411190].
  %%CITATION = EPHJA,C40,259;%%


\bibitem{SSA}R.~N.~Mohapatra and W.~Rodejohann,
  %``Scaling in the neutrino mass matrix,''
  {\it Phys.\ Lett.} {\bf B644}, 59 (2007) 
%  [arXiv:hep-ph/0608111].
  %%CITATION = PHLTA,B644,59;%%


 see also W.~Grimus and L.~Lavoura,
%   ``Softly broken lepton number L(e)-L(mu)-L(tau) with non-maximal solar
  %neutrino mixing,''
  {\it J.\ Phys.} {\bf G31}, 683 (2005)
%  [arXiv:hep-ph/0410279].
  %%CITATION = JPHGB,G31,683;%%

\bibitem{SSA2} A.~Blum, R.~N.~Mohapatra and W.~Rodejohann,
  %``Inverted Mass Hierarchy from Scaling in the Neutrino Mass Matrix: Low and
  %High Energy Phenomenology,''
  {\it Phys.\ Rev.} {\bf D76}, 053003 (2007)
%  [arXiv:0706.3801 [hep-ph]].
  %%CITATION = PHRVA,D76,053003;%%


\bibitem{MR}A.~Merle and W.~Rodejohann,
  %``The elements of the neutrino mass matrix: Allowed ranges and  implications
  %of texture zeros,''
  {\it Phys.\ Rev.}  {\bf D73}, 073012 (2006)
%  [arXiv:hep-ph/0603111].
  %%CITATION = PHRVA,D73,073012;%%

\bibitem{zeros}P.~H.~Frampton, S.~L.~Glashow and D.~Marfatia,
  %``Zeroes of the neutrino mass matrix,''
  {\it Phys.\ Lett.} {\bf B536}, 79 (2002)
%  [arXiv:hep-ph/0201008].
  %%CITATION = PHLTA,B536,79;%%


  Z.~z.~Xing,
  %``Texture zeros and Majorana phases of the neutrino mass matrix,''
  {\it Phys.\ Lett.} {\bf B530}, 159 (2002)
%  [arXiv:hep-ph/0201151].
  %%CITATION = PHLTA,B530,159;%%


  B.~R.~Desai, D.~P.~Roy and A.~R.~Vaucher,
  %``Three-neutrino mass matrices with two texture zeros,''
  {\it Mod.\ Phys.\ Lett.} {\bf A18}, 1355 (2003)
%  [arXiv:hep-ph/0209035].
  %%CITATION = MPLAE,A18,1355;%%


\bibitem{HR}C.~Hagedorn and W.~Rodejohann,
  %``Minimal mass matrices for Dirac neutrinos,''
  {\it JHEP} {\bf 0507}, 034 (2005)
%  [arXiv:hep-ph/0503143].
  %%CITATION = JHEPA,0507,034;%%

\bibitem{hybrid}S.~Kaneko, H.~Sawanaka and M.~Tanimoto,
  %``Hybrid textures of neutrinos,''
  {\it JHEP} {\bf 0508}, 073 (2005)
%  [arXiv:hep-ph/0504074].
  %%CITATION = JHEPA,0508,073;%%

\bibitem{det}G.~C.~Branco {\it et al.}, %R.~Gonzalez Felipe, F.~R.~Joaquim and T.~Yanagida,
  %``Removing ambiguities in the neutrino mass matrix,''
  {\it Phys.\ Lett.}  {\bf B562}, 265 (2003)
%  [arXiv:hep-ph/0212341].
  %%CITATION = PHLTA,B562,265;%%

\bibitem{trace}
D.~Black, A.~H.~Fariborz, S.~Nasri and J.~Schechter,
  %``Complementary ansatz for the neutrino mass matrix,''
  {\it Phys.\ Rev.} {\bf D62}, 073015 (2000)
%  [arXiv:hep-ph/0004105].
  %%CITATION = PHRVA,D62,073015;%%


  X.~G.~He and A.~Zee,
%   ``Neutrino masses with 'zero sum' condition: m(nu(1)) + m(nu(2)) +  m(nu(3))
  %= 0,''
  {\it Phys.\ Rev.} {\bf D68}, 037302 (2003)
%  [arXiv:hep-ph/0302201].
  %%CITATION = PHRVA,D68,037302;%%

  W.~Rodejohann,
  %``Neutrino mass matrices leaving no trace,''
  {\it Phys.\ Lett.} {\bf B579}, 127 (2004) 
%  [arXiv:hep-ph/0308119].
  %%CITATION = PHLTA,B579,127;%%


  S.~S.~Masood, S.~Nasri and J.~Schechter,
  %``Leptonic CP violation in a two parameter model,''
  {\it Phys.\ Rev.}  {\bf D71}, 093005 (2005)
%  [arXiv:hep-ph/0412401].
  %%CITATION = PHRVA,D71,093005;%%


\bibitem{carl0}
C.~H.~Albright and M.~C.~Chen,
  %``Model predictions for neutrino oscillation parameters,''
  Phys.\ Rev.\  D {\bf 74}, 113006 (2006)
%  [arXiv:hep-ph/0608137].
  %%CITATION = PHRVA,D74,113006;%%



\bibitem{I}P.~Minkowski, 
{\it Phys.\ Lett.} {\bf B67}, 421 (1977) 
%%CITATION = PHLTA,B67,421;%%

T.~Yanagida, \emph{Horizontal gauge symmetry and masses of neutrinos}, in
  \emph{Proceedings of the Workshop on The Unified Theory and the Baryon Number
  in the Universe} (O.~Sawada and A.~Sugamoto, eds.), KEK, Tsukuba, Japan,
  1979, p.~95 

S.~L. Glashow, \emph{The future of elementary particle physics}, in
  \emph{Proceedings of the 1979 Carg{\`e}se Summer Institute on Quarks and
  Leptons} (M.~L{\'e}vy, J.-L. Basdevant, D.~Speiser, 
J.~Weyers, R.~Gastmans, and M.~Jacob, eds.), 
Plenum Press, New York, 1980, p.~687 

M.~Gell-Mann, P.~Ramond, and R.~Slansky, \emph{Complex spinors and unified
  theories}, in \emph{Supergravity} (P.~van Nieuwenhuizen and D.~Z. Freedman,
  eds.), North Holland, Amsterdam, 1979%, p.~315 

R.~N. Mohapatra and G.~Senjanovi{\'c}, 
{\it Phys.\ Rev.\ Lett.} \textbf{44}, 912 (1980) 
%%CITATION = PRLTA,44,912;%%

\bibitem{revs}S.~F.~King,
  %``Neutrino mass models,''
  {\it Rept.\ Prog.\ Phys.} {\bf 67}, 107 (2004)
%  [arXiv:hep-ph/0310204].
  %%CITATION = RPPHA,67,107;%%

G.~Altarelli and F.~Feruglio,
  %``Models of neutrino masses and mixings,''
  {\it New J.\ Phys.}  {\bf 6}, 106 (2004)
%  [arXiv:hep-ph/0405048].
  %%CITATION = NJOPF,6,106;%%

G.~C.~Branco and M.~N.~Rebelo,
  %``Leptonic CP violation and neutrino mass models,''
  {\it New J.\ Phys.}  {\bf 7}, 86 (2005)
%  [arXiv:hep-ph/0411196].
  %%CITATION = NJOPF,7,86;%%

 R.~N.~Mohapatra,
  %``Seesaw mechanism and its implications,''
  arXiv:hep-ph/0412379 
  %%CITATION = HEP-PH/0412379;%%



\bibitem{planck}
R.~Barbieri, J.~R.~Ellis and M.~K.~Gaillard,
  %``Neutrino Masses And Oscillations In SU(5),''
  {\it Phys.\ Lett.}  {\bf B90}, 249 (1980) 
  %%CITATION = PHLTA,B90,249;%%


  E.~K.~Akhmedov, Z.~G.~Berezhiani and G.~Senjanovic,
  %``Planck scale physics and neutrino masses,''
  {\it Phys.\ Rev.\ Lett.} {\bf 69}, 3013 (1992)
%  [arXiv:hep-ph/9205230].
  %%CITATION = PRLTA,69,3013;%%

  E.~K.~Akhmedov, Z.~G.~Berezhiani, G.~Senjanovic and Z.~j.~Tao,
  %``Planck Scale Effects In Neutrino Physics,''
  {\it Phys.\ Rev.} {\bf D47}, 3245 (1993)
%  [arXiv:hep-ph/9208230].
  %%CITATION = PHRVA,D47,3245;%%

  A.~S.~Joshipura,
  %``Gravitationally violated U(1) symmetry and neutrino anomalies,''
  {\it Phys.\ Rev.} {\bf D60}, 053002 (1999)
%  [arXiv:hep-ph/9808261].
  %%CITATION = PHRVA,D60,053002;%%


  A.~de Gouvea and J.~W.~F.~Valle,
  %``Minimalistic neutrino mass model,''
  {\it Phys.\ Lett.} {\bf B501}, 115 (2001)
%  [arXiv:hep-ph/0010299].
  %%CITATION = PHLTA,B501,115;%%


  F.~Vissani, M.~Narayan and V.~Berezinsky,
  %``U(e3) from physics above the GUT scale,''
  {\it Phys.\ Lett.}  {\bf B571}, 209 (2003)
%  [arXiv:hep-ph/0305233].
  %%CITATION = PHLTA,B571,209;%%


%  B.~Singh Koranga, S.~U.~Sankar and M.~Narayan,
  %``Deviation from bimaximality due to Planck scale effects,''
%  arXiv:hep-ph/0611186 



 A.~Dighe, S.~Goswami and W.~Rodejohann,
  %``Corrections to Tri-bimaximal Neutrino Mixing: Renormalization and Planck
  %Scale Effects,''
  {\it Phys.\ Rev.}  {\bf D75}, 073023 (2007)
  %[arXiv:hep-ph/0612328].
  %%CITATION = PHRVA,D75,073023;%%

\bibitem{davidson} S.~Davidson and A.~Ibarra,
  %``Determining seesaw parameters from weak scale measurements?,''
  {\it JHEP} {\bf 0109}, 013 (2001)
%  [arXiv:hep-ph/0104076].
  %%CITATION = JHEPA,0109,013;%%



\bibitem{ellis}J.~R.~Ellis and M.~Raidal,
  %``Leptogenesis and the violation of lepton number and CP at low energies,''
  {\it Nucl.\ Phys.}  {\bf B643}, 229 (2002)
%  [arXiv:hep-ph/0206174].
  %%CITATION = NUPHA,B643,229;%%




\bibitem{PPR}S.~Pascoli, S.~T.~Petcov and W.~Rodejohann,
  %``On the connection of leptogenesis with low energy CP violation and LFV
  %charged lepton decays,''
  {\it Phys.\ Rev.}  {\bf D68}, 093007 (2003) 
%  [arXiv:hep-ph/0302054].
  %%CITATION = PHRVA,D68,093007;%%


\bibitem{branco}G.~C.~Branco {\it et al.}, %R.~Gonzalez Felipe, F.~R.~Joaquim and M.~N.~Rebelo,
  %``Leptogenesis, CP violation and neutrino data: What can we learn?,''
  {\it Nucl.\ Phys.}   {\bf B640}, 202 (2002)
%  [arXiv:hep-ph/0202030].
  %%CITATION = NUPHA,B640,202;%%


\bibitem{AFS}E.~K.~Akhmedov, M.~Frigerio and A.~Y.~Smirnov,
  %``Probing the seesaw mechanism with neutrino data and leptogenesis,''
  {\it JHEP} {\bf 0309}, 021 (2003)
%  [arXiv:hep-ph/0305322].
  %%CITATION = JHEPA,0309,021;%%


\bibitem{QLC0}
K.~A.~Hochmuth and W.~Rodejohann,
  %``Low and High Energy Phenomenology of Quark-Lepton Complementarity
  %Scenarios,''
  {\it Phys.\ Rev.} {\bf D75}, 073001 (2007)
%  [arXiv:hep-ph/0607103].
  %%CITATION = PHRVA,D75,073001;%%



\bibitem{rabi_chin}X.~d.~Ji, Y.~c.~Li, R.~N.~Mohapatra, 
S.~Nasri and Y.~Zhang,
  %``Leptogenesis in realistic SO(10) models,''
  {\it Phys.\ Lett.} {\bf B651}, 195 (2007) 
%  [arXiv:hep-ph/0605088].
  %%CITATION = PHLTA,B651,195;%%

\bibitem{carl} C.~H.~Albright and M.~C.~Chen,
  %``Lepton Flavor Violation in Predictive SUSY-GUT Models,''
  arXiv:0802.4228 [hep-ph]
  %%CITATION = ARXIV:0802.4228;%%

\bibitem{BPW} 
K.~S.~Babu, J.~C.~Pati and F.~Wilczek,
  %``Fermion masses, neutrino oscillations, and proton decay in the light of
  %SuperKamiokande,''
  {\it Nucl.\ Phys.} {\bf B566}, 33 (2000) 
%  [arXiv:hep-ph/9812538];
  %%CITATION = HEP-PH 9812538;%%

  K.~S.~Babu, J.~C.~Pati and P.~Rastogi,
  %``Tying in CP and flavor violations with fermion masses and neutrino
  %oscillations,''
  {\it Phys.\ Rev.} {\bf D71}, 015005 (2005) 
%  [arXiv:hep-ph/0410200];
  %%CITATION = HEP-PH 0410200;%%



\bibitem{GMN}  B.~Bajc, G.~Senjanovic and F.~Vissani,
{\it Phys.\ Rev.\ Lett.}  {\bf 90}, 051802 (2003) 
 %%CITATION = PRLTA,90,051802;%%


H.~S.~Goh, R.~N.~Mohapatra and S.~P.~Ng,
  %``Minimal SUSY SO(10), b tau unification and large neutrino mixings,''
  {\it Phys.\ Lett.} {\bf B570}, 215 (2003) and 
{\it Phys.\ Rev.} {\bf D68}, 115008 (2003)
%%CITATION = PHLTA,B570,215;%%
%%CITATION = PHRVA,D68,115008;%%


\bibitem{JLM}  X.~Ji, Y.~Li and R.~N.~Mohapatra,
  %``An SO(10) GUT model with lopsided mass matrix and neutrino mixing angle
  %theta(13),''
  {\it Phys.\ Lett.} {\bf B633}, 755 (2006) 
%%CITATION = PHLTA,B633,755;%%

\bibitem{DMM} B.~Dutta, Y.~Mimura and R.~N.~Mohapatra,
  {\it Phys.\ Rev.} {\bf D72}, 075009 (2005) 
%%CITATION = PHRVA,D72,075009;%%

\bibitem{AB} 
C.~H.~Albright and S.~M.~Barr,
  %``Realization of the large mixing angle solar neutrino solution in an
%SO(10)
  %supersymmetric grand unified model,''
  {\it Phys.\ Rev.} {\bf D64}, 073010 (2001) and 
%  [arXiv:hep-ph/0104294];
  %%CITATION = HEP-PH 0104294;%%
Phys.\ Rev.\  D {\bf 70}, 033013 (2004)
%  [arXiv:hep-ph/0404095].
  %%CITATION = PHRVA,D70,033013;%%



\bibitem{CI}J.~A.~Casas and A.~Ibarra,
  %``Oscillating neutrinos and mu --> e, gamma,''
  {\it Nucl.\ Phys.} {\bf B618}, 171 (2001)
%  [arXiv:hep-ph/0103065].
  %%CITATION = NUPHA,B618,171;%%


\bibitem{LFV}F.~Borzumati and A.~Masiero,
  %``Large Muon And Electron Number Violations In Supergravity Theories,''
  {\it Phys.\ Rev.\ Lett.} {\bf 57}, 961 (1986) 
  %%CITATION = PRLTA,57,961;%%

J.~Hisano, T.~Moroi, K.~Tobe and M.~Yamaguchi,
  %``Lepton-Flavor Violation via Right-Handed Neutrino Yukawa Couplings in
  %Supersymmetric Standard Model,''
  {\it Phys.\ Rev.} {\bf D53}, 2442 (1996) %[arXiv:hep-ph/9510309].
  %%CITATION = HEP-PH 9510309;%%


\bibitem{PDG}W.~M.~Yao {\it et al.}  [Particle Data Group],
  %``Review of particle physics,''
  {\it J.\ Phys.} {\bf G33}, 1 (2006)
  %%CITATION = JPHGB,G33,1;%%

\bibitem{mueg_lim}
M.~L.~Brooks {\it et al.}  [MEGA Collaboration],
  %``New limit for the family-number non-conserving decay mu+ --> e+ gamma,''
  {\it Phys.\ Rev.\ Lett.} {\bf 83}, 1521 (1999) %[arXiv:hep-ex/9905013].
  %%CITATION = PRLTA,83,1521;%%


\bibitem{teg_lim}B.~Aubert {\it et al.}  [BABAR Collaboration],
  %``Search for lepton flavor violation in the decay $\tau^\pm \to e^\pm
  %\gamma$,''
  {\it Phys.\ Rev.\ Lett.} {\bf 96}, 041801 (2006) %[arXiv:hep-ex/0508012].
  %%CITATION = PRLTA,96,041801;%%

\bibitem{tmg_lim}B.~Aubert {\it et al.}  [BABAR Collaboration],
  %``Search for lepton flavor violation in the decay $\tau \to \mu \gamma$,''
{\it Phys.\ Rev.\ Lett.} {\bf 95}, 041802 (2005) 
%[arXiv:hep-ex/0502032].
  %%CITATION = PRLTA,95,041802;%%


\bibitem{meg_fut}See the homepage of the MEG experiment, 
{\tt http://meg.web.psi.ch} 


\bibitem{BR_fut}A.~G.~Akeroyd {\it et al.}, hep-ex/0406071 
%%CITATION = HEP-EX 0406071;%%  

\bibitem{PPTS}
S.~T.~Petcov, S.~Profumo, Y.~Takanishi and C.~E.~Yaguna,
  %``Charged lepton flavor violating decays: Leading logarithmic  approximation
  %versus full RG results,''
  {\it Nucl.\ Phys.} {\bf B676}, 453 (2004) 
%  [arXiv:hep-ph/0306195].
  %%CITATION = NUPHA,B676,453;%%

\bibitem{mutauseesaw}
 R.~N.~Mohapatra and S.~Nasri,
  %``Leptogenesis and mu - tau symmetry,''
  {\it Phys.\ Rev.}  {\bf D71}, 033001 (2005)
 % [arXiv:hep-ph/0410369].
  %%CITATION = PHRVA,D71,033001;%%




\bibitem{QLC}
A.~Y.~Smirnov,
  %``Neutrinos: '...Annus mirabilis',''
  arXiv:hep-ph/0402264 
  %%CITATION = HEP-PH/0402264;%%

M.~Raidal,
  %``Prediction Theta(C) + Theta(sol) = pi/4 from flavor physics: A new
  %evidence for grand unification?,''
  {\it Phys.\ Rev.\ Lett.} {\bf 93}, 161801 (2004) 
%  [arXiv:hep-ph/0404046].
  %%CITATION = PRLTA,93,161801;%%

H.~Minakata and A.~Y.~Smirnov,
  %``Neutrino mixing and quark lepton complementarity,''
  {\it Phys.\ Rev.} {\bf D70}, 073009 (2004) 
%  [arXiv:hep-ph/0405088].
  %%CITATION = PHRVA,D70,073009;%%

K.~Cheung, S.~K.~Kang, C.~S.~Kim and J.~Lee,
  %``Lepton flavor violation as a probe of quark-lepton unification,''
  {\it Phys.\ Rev.} {\bf D72}, 036003 (2005) 
%  [arXiv:hep-ph/0503122].
  %%CITATION = PHRVA,D72,036003;%%


 M.~Picariello,
  %``Predictions for 'mu -> e gamma' in SUSY from non trivial Quark-Lepton
  %complementarity,''
  {\it Adv.\ High Energy Phys.} {\bf 2007}, 39676 (2007)
%  [arXiv:hep-ph/0703301].
  %%CITATION = 00642,2007,39676;%%

\bibitem{Wolf} L.~Wolfenstein,
  %``Parametrization Of The Kobayashi-Maskawa Matrix,''
  {\it Phys.\ Rev.\ Lett.} {\bf 51}, 1945 (1983) 
  %%CITATION = PRLTA,51,1945;%%



\bibitem{CM}M.~C.~J.~Chen and K.~T.~Mahanthappa,
  %``Lepton flavor violating decays, soft leptogenesis and SUSY SO(10),''
  {\it Phys.\ Rev.}  {\bf D70}, 113013 (2004)
%  [arXiv:hep-ph/0409096].
  %%CITATION = PHRVA,D70,113013;%%

\bibitem{CY}Y.~Cai and H.~B.~Yu,
  %``A SO(10) GUT model with S4 flavor symmetry,''
  {\it Phys.\ Rev.} {\bf D74}, 115005 (2006)
%  [arXiv:hep-ph/0608022].
  %%CITATION = PHRVA,D74,115005;%%

\bibitem{DR} R.~Dermisek and S.~Raby,
%   ``Bi-large neutrino mixing and CP violation in an SO(10) SUSY GUT for
  %fermion masses,''
  {\it Phys.\ Lett.} {\bf B622}, 327 (2005)
%  [arXiv:hep-ph/0507045].
  %%CITATION = PHLTA,B622,327;%%

\bibitem{GK}W.~Grimus and H.~Kuhbock,
  %``Fermion masses and mixings in a renormalizable SO(10) x Z(2) GUT,''
  {\it Phys.\ Lett.}  {\bf B643}, 182 (2006)
%  [arXiv:hep-ph/0607197].
  %%CITATION = PHLTA,B643,182;%%

\bibitem{lepto} M.~Fukugita and T.~Yanagida,
  %``Baryogenesis Without Grand Unification,''
  {\it Phys.\ Lett.}  {\bf B174}, 45 (1986) 
  %%CITATION = PHLTA,B174,45;%%

\bibitem{D}For a recent review, see 
S.~Davidson, E.~Nardi and Y.~Nir,
  %``Leptogenesis,''
  arXiv:0802.2962 [hep-ph] 
  %%CITATION = ARXIV:0802.2962;

\bibitem{flavor_flav}
A.~Abada {\it et al.}, %S.~Davidson, F.~X.~Josse-Michaux, M.~Losada and A.~Riotto,
  %``Flavour issues in leptogenesis,''
  {\it JCAP} {\bf 0604}, 004 (2006) %[arXiv:hep-ph/0601083]; 

  %%CITATION = JCAPA,0604,004;%%
 E.~Nardi, Y.~Nir, E.~Roulet and J.~Racker,
  %``The importance of flavor in leptogenesis,''
  {\it JHEP} {\bf 0601}, 164 (2006) %; [arXiv:hep-ph/0601084]; 
  %%CITATION = JHEPA,0601,164;%%

\bibitem{flavor_others}
F.~X.~Josse-Michaux and A.~Abada,
  %``Study of flavour dependencies in leptogenesis,''
  {\it JCAP} {\bf 0710}, 009 (2007)
%  [arXiv:hep-ph/0703084].
  %%CITATION = JCAPA,0710,009;%%

S.~Antusch, S.~F.~King and A.~Riotto,
  %``Flavour-dependent leptogenesis with sequential dominance,''
  {\it JCAP} {\bf 0611}, 011 (2006)
%  [arXiv:hep-ph/0609038].
  %%CITATION = JCAPA,0611,011;%%

recent overviews are %given in 
 S.~Blanchet and P.~Di Bari,
  %``Flavor effects in thermal leptogenesis,''
  {\it Nucl.\ Phys.\ Proc.\ Suppl.}  {\bf 168}, 372 (2007) %[hep-ph/0702089]; 
  %%CITATION = NUPHZ,168,372;%%

S.~Davidson,
  %``Flavoured Leptogenesis,''
  arXiv:0705.1590 [hep-ph] 
  %%CITATION = ARXIV:0705.1590;%%

%\bibitem{flavor_old} 

see also 
R.~Barbieri, P.~Creminelli, A.~Strumia and N.~Tetradis,
  %``Baryogenesis through leptogenesis,''
  {\it Nucl.\ Phys.} {\bf B575}, 61 (2000) %; [arXiv:hep-ph/9911315]; 
  %%CITATION = NUPHA,B575,61;%%

T.~Endoh, T.~Morozumi and Z.~h.~Xiong,
  %``Primordial lepton family asymmetries in seesaw model,''
  {\it Prog.\ Theor.\ Phys.}  {\bf 111}, 123 (2004)

%  [arXiv:hep-ph/0308276].
  %%CITATION = PTPKA,111,123;%%



\bibitem{Bari} S.~Blanchet and P.~Di Bari,
  %``Flavor effects on leptogenesis predictions,''
  {\it JCAP} {\bf 0703}, 018 (2007)
%  [arXiv:hep-ph/0607330].
  %%CITATION = JCAPA,0703,018;%%

 A.~Anisimov, S.~Blanchet and P.~Di Bari,
  %``Viability of Dirac phase leptogenesis,''
  arXiv:0707.3024 [hep-ph]
  %%CITATION = ARXIV:0707.3024;%%


\bibitem{petcov_flav}S.~Pascoli, S.~T.~Petcov and A.~Riotto,
  %``Connecting low energy leptonic CP-violation to leptogenesis,''
  {\it Phys.\ Rev.} {\bf D75}, 083511 (2007) and 
{\it Nucl.\ Phys.} {\bf B774}, 1 (2007) 
%  [arXiv:hep-ph/0609125].
  %%CITATION = PHRVA,D75,083511;%%

%S.~Pascoli, S.~T.~Petcov and A.~Riotto,
  %``Leptogenesis and low energy CP violation in neutrino physics,''
   
%  [arXiv:hep-ph/0611338].
  %%CITATION = NUPHA,B774,1;%%

E.~Molinaro and S.~T.~Petcov,
  %``The Interplay Between the 'Low' and 'High' Energy CP-Violation in
  %Leptogenesis,''
  arXiv:0803.4120 [hep-ph] 
  %%CITATION = ARXIV:0803.4120;%%


\bibitem{br}G.~C.~Branco, R.~Gonzalez Felipe and F.~R.~Joaquim,
  %``A new bridge between leptonic CP violation and leptogenesis,''
  {\it Phys.\ Lett.}  {\bf B645}, 432 (2007)
%  [arXiv:hep-ph/0609297].
  %%CITATION = PHLTA,B645,432;%%

\bibitem{sasha}S.~Davidson, J.~Garayoa, F.~Palorini and N.~Rius,
  %``Insensitivity of flavoured leptogenesis to low energy CP violation,''
  {\it Phys.\ Rev.\ Lett.} {\bf 99}, 161801 (2007)
%  [arXiv:0705.1503 [hep-ph]].
  %%CITATION = PRLTA,99,161801;%%



\bibitem{wmap}J.~Dunkley {\it et al.}  [WMAP Collaboration],
  %``Five-Year Wilkinson Microwave Anisotropy Probe (WMAP) Observations:
  %Likelihoods and Parameters from the WMAP data,''
  arXiv:0803.0586 [astro-ph]
  %%CITATION = ARXIV:0803.0586;%%

\bibitem{nocp}G.~C.~Branco, T.~Morozumi, B.~M.~Nobre and M.~N.~Rebelo,
  %``A bridge between CP violation at low energies and leptogenesis,''
  {\it Nucl.\ Phys.} {\bf B617}, 475 (2001)
%  [arXiv:hep-ph/0107164].
  %%CITATION = NUPHA,B617,475;%%




\bibitem{resyb}
M.~Flanz, E.~A.~Paschos and U.~Sarkar,
  %``Baryogenesis from a lepton asymmetric universe,''
  {\it Phys.\ Lett.} {\bf B345}, 248 (1995)
%  [Erratum-ibid.\  {\bf B382}, 447 (1996)]
%  [arXiv:hep-ph/9411366].
  %%CITATION = PHLTA,B345,248;%%


  M.~Flanz, E.~A.~Paschos, U.~Sarkar and J.~Weiss,
  %``Baryogenesis through mixing of heavy Majorana neutrinos,''
  {\it Phys.\ Lett.} {\bf B389}, 693 (1996)
%  [arXiv:hep-ph/9607310].
  %%CITATION = PHLTA,B389,693;%%


  L.~Covi, E.~Roulet and F.~Vissani,
  %``CP violating decays in leptogenesis scenarios,''
  {\it Phys.\ Lett.} {\bf B384}, 169 (1996)
%  [arXiv:hep-ph/9605319].
  %%CITATION = PHLTA,B384,169;%%


  A.~Pilaftsis,
  %``CP violation and baryogenesis due to heavy Majorana neutrinos,''
  {\it Phys.\ Rev.} {\bf D56}, 5431 (1997)
%  [arXiv:hep-ph/9707235].
  %%CITATION = PHRVA,D56,5431;%%

A.~Pilaftsis and T.~E.~J.~Underwood,
  %``Resonant leptogenesis,''
  {\it Nucl.\ Phys.} {\bf B692}, 303 (2004)
%  [arXiv:hep-ph/0309342].
  %%CITATION = NUPHA,B692,303;%%




\bibitem{mMold}See e.g.~ W.~Buchm\"uller, P.~Di Bari and M.~Pl\"umacher,
  %``Leptogenesis for pedestrians,''
  {\it Annals Phys.} {\bf 315}, 305 (2005)
%  [arXiv:hep-ph/0401240].
  %%CITATION = APNYA,315,305;%%


 G.~F.~Giudice, A.~Notari, M.~Raidal, A.~Riotto and A.~Strumia,
  %``Towards a complete theory of thermal leptogenesis in the SM and MSSM,''
  {\it Nucl.\ Phys.} {\bf B685}, 89 (2004) 

%  [arXiv:hep-ph/0310123].
  %%CITATION = NUPHA,B685,89;%%



\bibitem{DI}S.~Davidson and A.~Ibarra,
  %``A lower bound on the right-handed neutrino mass from leptogenesis,''
  {\it Phys.\ Lett.}  {\bf B535}, 25 (2002)
%  [arXiv:hep-ph/0202239].
  %%CITATION = PHLTA,B535,25;%%

\bibitem{comb}See e.g.~S.~Lavignac, I.~Masina and C.~A.~Savoy,
  %``tau --> mu gamma and mu --> e gamma as probes of neutrino mass models,''
  {\it Phys.\ Lett.}   {\bf B520}, 269 (2001)
%  [arXiv:hep-ph/0106245].
  %%CITATION = PHLTA,B520,269;%%

J.~R.~Ellis, J.~Hisano, S.~Lola and M.~Raidal,
  %``CP violation in the minimal supersymmetric seesaw model,''
  {\it Nucl.\ Phys.} {\bf B621}, 208 (2002)
%  [arXiv:hep-ph/0109125].
  %%CITATION = NUPHA,B621,208;%%


F.~Deppisch {\it et al.}, %H.~P\"as, A.~Redelbach, R.~R\"uckl and Y.~Shimizu,
  %``Probing the Majorana mass scale of right-handed neutrinos in mSUGRA,''
  {\it Eur.\ Phys.\ J.}  {\bf C28}, 365 (2003) and 
 {\it Phys.\ Rev.} {\bf D73}, 033004 (2006)
%  [arXiv:hep-ph/0511062].
  %%CITATION = PHRVA,D73,033004;%%
%  [arXiv:hep-ph/0206122].
  %%CITATION = EPHJA,C28,365;%%


  A.~Masiero, S.~K.~Vempati and O.~Vives,
  %``Seesaw and lepton flavour violation in SUSY SO(10),''
  {\it Nucl.\ Phys.} {\bf B649}, 189 (2003)
%  [arXiv:hep-ph/0209303].
  %%CITATION = NUPHA,B649,189;%%

  S.~Pascoli, S.~T.~Petcov and C.~E.~Yaguna,
%   ``Quasi-degenerate neutrino mass spectrum, mu --> e + gamma decay and
  %leptogenesis,''
  {\it Phys.\ Lett.} {\bf B564}, 241 (2003)
%  [arXiv:hep-ph/0301095].
  %%CITATION = PHLTA,B564,241;%%

%S.~Kanemura {\it et al.}, %K.~Matsuda, T.~Ota, T.~Shindou, E.~Takasugi and K.~Tsumura,
%   ``Enhancement of lepton flavor violation in a model with bi-maximal  mixing
  %at the grand unification scale,''
%  {\it Phys.\ Rev.} {\bf D72}, 093004 (2005)
%  [arXiv:hep-ph/0501228].



S.~Antusch, E.~Arganda, M.~J.~Herrero and A.~M.~Teixeira,
%   ``Impact of theta(13) on lepton flavour violating processes within SUSY
  %seesaw,''
  {\it JHEP} {\bf 0611}, 090 (2006)
%  [arXiv:hep-ph/0607263].
  %%CITATION = JHEPA,0611,090;%%


  G.~C.~Branco {\it et al.}, %A.~J.~Buras, S.~Jager, S.~Uhlig and A.~Weiler,
%   ``Another look at minimal lepton flavour violation, l(i) --> l(j) gamma,
  %leptogenesis, and the ratio M(nu)/Lambda(LFV),''
  {\it JHEP} {\bf 0709}, 004 (2007)
%  [arXiv:hep-ph/0609067].
  %%CITATION = JHEPA,0709,004;%%

S.~Antusch and A.~M.~Teixeira,
  %``Towards constraints on the SUSY seesaw from flavour-dependent
  %leptogenesis,''
  {\it JCAP} {\bf 0702}, 024 (2007)
%  [arXiv:hep-ph/0611232].
  %%CITATION = JCAPA,0702,024;%%

  J.~A.~Casas, A.~Ibarra and F.~Jimenez-Alburquerque,
%   ``Hints on the high-energy seesaw mechanism from the low-energy neutrino
  %spectrum,''
  {\it JHEP} {\bf 0704}, 064 (2007) 
%  [arXiv:hep-ph/0612289].
  %%CITATION = JHEPA,0704,064;%%

E.~J.~Chun, J.~L.~Evans, D.~E.~Morrissey and J.~D.~Wells,
  %``Higgs Boson Exempt No-Scale Supersymmetry with a Neutrino Seesaw:
  %Implications for Lepton Flavor Violation and Leptogenesis,''
  arXiv:0804.3050 [hep-ph] 
  %%CITATION = ARXIV:0804.3050;%%


\bibitem{PRST} S.~T.~Petcov, W.~Rodejohann, T.~Shindou and Y.~Takanishi,
  %``The see-saw mechanism, neutrino Yukawa couplings, LFV decays l(i) -->  l(j)
  %+ gamma and leptogenesis,''
  {\it Nucl.\ Phys.} {\bf B739}, 208 (2006)
%  [arXiv:hep-ph/0510404].
  %%CITATION = NUPHA,B739,208;%%

\bibitem{zero}A.~Ibarra and C.~Simonetto,
  %``Constraints on the rare tau decays from mu --> e gamma in the
  %supersymmetric see-saw model,''
  arXiv:0802.3858 [hep-ph] 
  %%CITATION = ARXIV:0802.3858;%%


\bibitem{sing}K.~L.~McDonald and B.~H.~J.~McKellar,
  %``The non-canonical singular see-saw mechanism,''
  hep-ph/0401073 
  %%CITATION = HEP-PH 0401073;%%

G.~J.~Stephenson {\it et al.}, 
%, T.~Goldman, B.~H.~J.~McKellar and M.~Garbutt,
  %``3+2 neutrinos in a see-saw variation,''
  hep-ph/0307245
  %%CITATION = HEP-PH 0307245;%%

\bibitem{III}
R.~Foot, H.~Lew, X.~G.~He and G.~C.~Joshi,
  %``SEESAW NEUTRINO MASSES INDUCED BY A TRIPLET OF LEPTONS,''
  {\it Z.\ Phys.} {\bf C44}, 441 (1989) 
  %%CITATION = ZEPYA,C44,441;%%

E.~Ma,
  %``Pathways to naturally small neutrino masses,''
  {\it Phys.\ Rev.\ Lett.} {\bf 81}, 1171 (1998) 
%  [arXiv:hep-ph/9805219].
  %%CITATION = PRLTA,81,1171;%%


\bibitem{cascade}R.~N.~Mohapatra,
  %``MECHANISM FOR UNDERSTANDING SMALL NEUTRINO MASS IN SUPERSTRING THEORIES,''
  {\it Phys.\ Rev.\ Lett.} {\bf 56}, 561 (1986) 
  %%CITATION = PRLTA,56,561;%%


  R.~N.~Mohapatra and J.~W.~F.~Valle,
  %``NEUTRINO MASS AND BARYON-NUMBER NONCONSERVATION IN SUPERSTRING MODELS,''
  {\it Phys.\ Rev.}  {\bf D34}, 1642 (1986) 
  %%CITATION = PHRVA,D34,1642;%%


\bibitem{screening}M.~Lindner, M.~A.~Schmidt and A.~Y.~Smirnov,
  %``Screening of Dirac flavor structure in the seesaw and neutrino mixing,''
  {\it JHEP} {\bf 0507}, 048 (2005)

%  [arXiv:hep-ph/0505067].
  %%CITATION = JHEPA,0507,048;%%


\bibitem{barr}E.~K.~Akhmedov, M.~Lindner, E.~Schnapka and J.~W.~F.~Valle,
  %``Left-Right Symmetry Breaking In Njl Approach,''
  {\it Phys.\ Lett.} {\bf B368}, 270 (1996) and 
{\it Phys.\ Rev.} {\bf D53}, 2752 (1996)
%  [arXiv:hep-ph/9507275].
  %%CITATION = PHLTA,B368,270;%%
  %%CITATION = PHRVA,D53,2752;%%

S.~M.~Barr,
  %``A different see-saw formula for neutrino masses,''
  Phys.\ Rev.\ Lett.\  {\bf 92}, 101601 (2004)
%  [arXiv:hep-ph/0309152].
  %%CITATION = PRLTA,92,101601;%%

M.~Malinsky, J.~C.~Romao and J.~W.~F.~Valle,
  %``Novel supersymmetric SO(10) seesaw mechanism,''
  {\it Phys.\ Rev.\ Lett.} {\bf 95}, 161801 (2005)

%  [arXiv:hep-ph/0506296].
  %%CITATION = PRLTA,95,161801;%%

\bibitem{seesaw?}
A.~Y.~Smirnov,
  %``Alternatives to the seesaw mechanism,''
  arXiv:hep-ph/0411194.
  %%CITATION = HEP-PH/0411194;%%



\bibitem{II}M.~Magg and C.~Wetterich,
%``Neutrino Mass Problem And Gauge Hierarchy,''
{\it Phys.\ Lett.} {\bf B94}, 61 (1980) 
%%CITATION = PHLTA,B94,61;%%

R.~N.~Mohapatra and G.~Senjanovic,
% ``Neutrino Masses And Mixings In Gauge Models With Spontaneous Parity
%Violation,''
%
{\it Phys.\ Rev.} {\bf D23}, 165 (1981) 
%%CITATION = PHRVA,D23,165;%%

G.~Lazarides, Q.~Shafi and C.~Wetterich,
%``Proton Lifetime And Fermion Masses In An SO(10) Model,''
{\it Nucl.\ Phys.} {\bf B181}, 287 (1981) 
%%CITATION = NUPHA,B181,287;%%

J.~Schechter and J.~W.~F.~Valle,
%``Neutrino Masses In SU(2) X U(1) Theories,''
{\it Phys.\ Rev.} {\bf D22}, 2227 (1980) 
%%CITATION = PHRVA,D22,2227;%%


\bibitem{lepto_II}P.~J.~O'Donnell and U.~Sarkar,
  %``Baryogenesis via lepton number violating scalar interactions,''
  {\it Phys.\ Rev.}  {\bf D49}, 2118 (1994)
  [arXiv:hep-ph/9307279].
  %%CITATION = PHRVA,D49,2118;%%

%\cite{Ma:1998dx}
%\bibitem{Ma:1998dx}
  E.~Ma and U.~Sarkar,
  %``Neutrino masses and leptogenesis with heavy Higgs triplets,''
  {\it Phys.\ Rev.\ Lett.} {\bf 80}, 5716 (1998)
%  [arXiv:hep-ph/9802445].
  %%CITATION = PRLTA,80,5716;%%

%\cite{Hambye:2000ui}
%\bibitem{Hambye:2000ui}
  T.~Hambye, E.~Ma and U.~Sarkar,
  %``Supersymmetric triplet Higgs model of neutrino masses and leptogenesis,''
  {\it Nucl.\ Phys.}   {\bf B602}, 23 (2001)
%  [arXiv:hep-ph/0011192].
  %%CITATION = NUPHA,B602,23;%%




 

\bibitem{Rossi0}
A.~Rossi,
  %``Supersymmetric seesaw without singlet neutrinos: Neutrino masses and
  %lepton-flavour violation,''
  {\it Phys.\ Rev.} {\bf D 66}, 075003 (2002) 
%  [arXiv:hep-ph/0207006].
  %%CITATION = PHRVA,D66,075003;%%

E.~J.~Chun, A.~Masiero, A.~Rossi and S.~K.~Vempati,
  %``A predictive seesaw scenario for EDMs,''
  {\it Phys.\ Lett.}  {\bf B 622}, 112 (2005) 
%  [arXiv:hep-ph/0502022].
  %%CITATION = PHLTA,B622,112;%%

F.~R.~Joaquim and A.~Rossi,
  %``Gauge and Yukawa mediated supersymmetry breaking in the triplet seesaw
  %scenario,''
  {\it Phys.\ Rev.\ Lett.}  {\bf 97}, 181801 (2006)  
%  [arXiv:hep-ph/0604083].
  %%CITATION = PRLTA,97,181801;%%



\bibitem{IILFV}
E.~J.~Chun, K.~Y.~Lee and S.~C.~Park,
  %``Testing Higgs triplet model and neutrino mass patterns,''
  {\it Phys.\ Lett. }  {\bf B566}, 142 (2003) 
%  [arXiv:hep-ph/0304069].
  %%CITATION = PHLTA,B566,142;%%

M.~Kakizaki, Y.~Ogura and F.~Shima,
  %``Lepton flavor violation in the triplet Higgs model,''
  {\it Phys.\ Lett.}  {\bf B566}, 210 (2003) 
%  [arXiv:hep-ph/0304254].
  %%CITATION = PHLTA,B566,210;%%



\bibitem{Rossi1}F.~R.~Joaquim and A.~Rossi,
  %``Phenomenology of the triplet seesaw mechanism with gauge and Yukawa
  %mediation of SUSY breaking,''
  {\it Nucl.\ Phys.} {\bf B765}, 71 (2007) 
%  [arXiv:hep-ph/0607298].
  %%CITATION = NUPHA,B765,71;%%


\bibitem{h120}Y.~Farzan and A.~Y.~Smirnov,
  %``Leptonic CP violation: Zero, maximal or between the two extremes,''
  {\it JHEP} {\bf 0701}, 059 (2007)
%  [arXiv:hep-ph/0610337].
  %%CITATION = JHEPA,0701,059;%%



\bibitem{IIYB}
T.~Hambye and G.~Senjanovi{\'c},
  %``Consequences of triplet seesaw for leptogenesis,''
  {\it Phys.\ Lett.} {\bf B582}, 73 (2004) 
%  [arXiv:hep-ph/0307237].
  %%CITATION = PHLTA,B582,73;%%

 S.~Antusch and S.~F.~King,
  %``Type II leptogenesis and the neutrino mass scale,''
  {\it Phys.\ Lett.} {\bf B597}, 199 (2004)  
%  [arXiv:hep-ph/0405093].
  %%CITATION = PHLTA,B597,199;%%

S.~Antusch,
  %``Flavour-dependent type II leptogenesis,''
  {\it Phys.\ Rev.}  {\bf D76}, 023512 (2007) 
%  [arXiv:0704.1591 [hep-ph]].
  %%CITATION = PHRVA,D76,023512;%%

\bibitem{LR}
M.~Lindner and W.~Rodejohann,
  %``Large and Almost Maximal Neutrino Mixing within the Type II See-Saw
  %Mechanism,''
  {\it JHEP} {\bf 0705}, 089 (2007)
%  [arXiv:hep-ph/0703171].
  %%CITATION = JHEPA,0705,089;%%

\bibitem{AR}E.~K.~Akhmedov and W.~Rodejohann,
  %``A Yukawa coupling parameterization for type I+ II seesaw formula and
  %applications to lepton flavor violation and leptogenesis,''
  arXiv:0803.2417 [hep-ph] 
  %%CITATION = ARXIV:0803.2417;%%



\end{thebibliography}
\end{document}